\documentclass[letterpaper,12pt]{article}
\usepackage{morefloats}
\usepackage{amsmath}
\usepackage{amssymb}
\usepackage[dvips]{graphicx,epsfig}
\usepackage{psfrag}
\title{Electromagnetic quasinormal modes of $D$-dimensional black holes II}

\author{A.\ L\'opez-Ortega\thanks{Electronic address: alopezortega@ucol.mx and alopezortega@gmail.com}  \\ 
Facultad de Ciencias \\ 
Universidad de Colima \\ 
Bernal Diaz del Castillo No. 340  \\ 
Colima, Colima, Mexico\\ } 

\begin{document}

\maketitle

\begin{abstract}

By using the sixth order WKB approximation we calculate for an electromagnetic field propagating in $D$-dimensional Schwarzschild and Schwarzschild de Sitter (SdS) black holes its quasinormal (QN) frequencies for the fundamental mode and first overtones. We study the dependence of these QN frequencies on the value of the cosmological constant and the spacetime dimension. We also compare with the known results for the gravitational perturbations propagating in the same background. Moreover we exactly compute the QN frequencies of the electromagnetic field propagating in  $D$-dimen\-sional massless topological black hole and for charged $D$-dimen\-sional Nariai spacetime we exactly calculate the QN frequencies of the coupled electromagnetic and gravitational perturbations.

\end{abstract}



\textbf{Running title}\,\,\,  de Sitter quasinormal modes

\section{Introduction}

The quasinormal modes (QNMs) are characteristic oscillations of a black hole that depend on its parameters, for example, the mass, electric charge, and angular momentum. For an asymptotically flat black hole these modes are purely ingoing near the event horizon and purely outgoing at infinity. For other black holes the imposed condition far from the horizon can be different \cite{Kokkotas:1999bd}, for example, we can impose that the field vanishes at infinity for asymptotically anti-de Sitter black holes \cite{Kokkotas:1999bd}, \cite{Horowitz:1999jd}, although other boundary conditions may be used.

In four-dimensional asymptotically flat backgrounds the QNMs of gravitational perturbations are studied to explore the linear stability of the black holes and because it is expected that these modes may be detected by the gravitational wave observatories \cite{Kokkotas:1999bd}. Moreover, in recent times the QNMs have been useful in other research lines \cite{Kokkotas:1999bd}--\cite{Hod:1998vk}.

Recently the computation of the quasinormal (QN) frequencies of several higher dimensional black holes has attracted much attention. This interest is mainly motivated by the possibles applications of the QN frequencies in the following research lines a) the AdS-CFT correspondence of String Theory \cite{Horowitz:1999jd}, b)  the study of the higher dimensional features of General Relativity \cite{Bizon:2005cp}, c) the analysis of the physical implications of the different Brane World models \cite{Kanti:2004nr}, and d) the understanding of the thermodynamical properties of black holes in Loop Quantum Gravity \cite{Hod:1998vk}.

Therefore there are many references where the QN frequencies of different higher dimensional backgrounds have been calculated, mainly for the Klein-Gordon and gravitational perturbations (for some examples see Refs.\ \cite{Motl:2003cd}--\cite{Lopez-Ortega:2006my}). In several of these references the analytical values of the asymptotic QN frequencies have been found \cite{Motl:2003cd}--\cite{Lopez-Ortega:2006vn}. In other the QN frequencies for low multipole number $l$ and mode number $n$ were computed by using numerical and semi-analytical methods \cite{Cardoso:2003vt}--\cite{Konoplya:2003dd}. Also, there are some higher dimensional backgrounds for which their QN frequencies have been exactly computed \cite{Birmingham:2006zx}--\cite{Lopez-Ortega:2006my}.

As is well known, sometimes the electromagnetic field behaves in a different way than the Klein-Gordon and gravitational perturbations. For example in a recent paper \cite{Lopez-Ortega:2006vn} (see also \cite{Motl:2003cd}), we calculate the asymptotic QN frequencies of the electromagnetic field propagating in $D$-dimensional Schwarzschild, Schwarzschild de Sitter (SdS), and Schwarzschild anti-de Sitter black holes. In that reference we find that there are some differences in the asymptotic behavior of the QN frequencies for the electromagnetic and gravitational perturbations. Thus we believe that the computation of the QN frequencies for the electromagnetic field propagating in $D$-dimensional spherically symmetric backgrounds is an interesting problem.

As far as we know, the low mode number QN frequencies of the electromagnetic field of vector type have been computed for Schwarzschild spacetime in five dimensions by Cardoso, et al.\ in Ref.\ \cite{Cardoso:2003vt} and some new results appear in Ref.\ \cite{Konoplya:2007jv}, too for the vector type electromagnetic perturbation. For  $D$-dimensional SdS black holes only recently were published some results for the QN frequencies of the electromagnetic field \cite{Konoplya:2007jv}. Here we extend the findings of these references, since for both vector type and scalar type  electromagnetic perturbations propagating in Schwarzschild and SdS backgrounds, in various spacetime dimensions, we calculate and study in detail the QN frequencies of the fundamental mode and first overtones for a larger range of multipole numbers.

Thus using the WKB method proposed and first employed in Refs.\ \cite{Schutz-Will:1985}--\cite{Kokkotas:1988fm}, we compute the low mode number QN frequencies of the electromagnetic field propagating in $D$-dimensional Schwarzschild spacetime for $D=5,6,7,8,9,10$. For SdS background we also calculate some low mode number QN frequencies in $D=5,6,7,8$. For both spacetimes we compare with the previous results for the gravitational perturbations moving in the same background. We also find the large angular momentum limit of the QN frequencies for the electromagnetic field propagating in $D$-dimensional SdS spacetime. 

Moreover, following Birmingham and Mokhtari \cite{Birmingham:2006zx} we exactly calculate the QN frequencies of the electromagnetic field propagating in a $D$-dimensional massless topological black hole. This spacetime can be considered as a $D$-dimensional generalization of the  three-dimensional BTZ black hole \cite{Birmingham:2006zx}. Finally, for $D$-dimensional charged Nariai spacetimes we also calculate analytically the QN frequencies of the coupled electromagnetic and gravitational perturbations. 

This paper is organized as follows. In Sect.\ \ref{Section EM}, following Kodama and Ishibashi \cite{Kodama:2003kk}, we write the effective potentials of the Schr\"odinger type equations corresponding to the two types of electromagnetic perturbations propagating in $D$-di\-men\-sion\-al black hole whose event horizon is a $p$-dimensional Einstein manifold ($p=D-2$). In Sect.\ \ref{Section WKB} we explain the basic facts on the WKB approximation that we use in the following two sections to compute the QN frequencies of the electromagnetic field. In Sect.\ \ref{Section Schwarzschild} we calculate the low mode number QN frequencies of the electromagnetic field propagating in $D$-dimensional Schwarzschild black hole for $D=5,6,7,8,9,10$. For both types of electromagnetic perturbations we compare their oscillation frequencies and damping rates and we also compare these quantities of the electromagnetic fields with those of the gravitational perturbations. Also, for fixed angular momentum and mode numbers, we briefly study the dependence of the QN frequencies on the dimension of Schwarzschild spacetime. 

As in Sect.\ \ref{Section Schwarzschild} for $D$-dimensional Schwarzschild background, in Sect.\ \ref{Section SdS} we compute the QN frequencies of the electromagnetic field moving in $D$-dimen\-sion\-al SdS black hole for $D=5,6,7,8$. In particular, for fixed angular momentum and mode numbers, we study in detail the dependence of its QN frequencies on the value of the cosmological constant and on the spacetime dimension. In Subsect.\ \ref{Subsection Large l} we find the large angular momentum limit of the QN frequencies for the electromagnetic field propagating in $D$-dimensional SdS black hole. In Sect.\ \ref{Section Birmingham} we exactly calculate the QN frequencies of the electromagnetic field propagating in a massless topological black hole. In Sect.\ \ref{Discussion section} we discuss the main results of this paper. Finally in Appendix \ref{Appendix Nariai}, we exactly compute the QN frequencies of the coupled electromagnetic and gravitational perturbations propagating in $D$-dimensional charged Nariai background.

\section{Electromagnetic field}
\label{Section EM}

If we take $f$ as a function of the radial coordinate $r$ and ${\rm d} \sigma^2_p$ as the line element of a $p$-dimensional Einstein space, then the metric of the form
\begin{equation} \label{e: metric general}
{\rm d} s^2 = g_{\alpha \beta} x^\alpha x^\beta = -f(r)\, {\rm d}t^2 + \frac{ {\rm d}r^2}{f(r)} + r^2 {\rm d} \sigma^2_p, \qquad \alpha,\,\beta=t,r,x^1,\dots,x^p ,
\end{equation} 
includes many solutions of the $D$-dimensional Einstein-Maxwell equations with cosmological constant ($D=p+2$) \cite{Kodama:2003kk}--\cite{b: Nariai solution}. In Ref.\ \cite{Kodama:2003kk} Kodama and Ishibashi showed that when the spacetime has a metric of the form (\ref{e: metric general}), for each perturbation type (scalar, vector, or tensor), the equations of motion for the coupled electromagnetic and gravitational perturbations simplify to Schr\"odinger type equations\footnote{See also Ref.\ \cite{Kodama:2003jz}.}
\begin{equation} \label{eq: Schrodinger type equation}
\frac{{\rm d}^2 \Phi }{{\rm d}x^2}  + (\omega^2 - V(r)) \Phi = 0,
\end{equation} 
where $x$ is the tortoise coordinate defined by
\begin{equation} \label{e: tortoise coordinate}
 x =  \int \frac{{\rm d} r}{f(r)}, 
\end{equation} 
and the effective potentials $V(r)$ are complicated functions of the coordinate $r$ and the black hole parameters (for example see formulas (3.7), (4.38), and (5.61) in Ref.\ \cite{Kodama:2003kk}).

In the main body of this paper we study the QNMs of the electromagnetic field propagating in Schwarzschild, SdS, and massless topological black holes. So we use the equations of Ref.\ \cite{Kodama:2003kk} when the electric charge is equal to zero, that is, the electromagnetic and gravitational perturbations are decoupled.\footnote{In Appendix \ref{Appendix Nariai} we use a special case of the equations for the coupled electromagnetic and gravitational perturbations of Ref.\ \cite{Kodama:2003kk}.}  

For this case the equations of motion for the electromagnetic perturbations simplify to Schr\"odinger type equations (\ref{eq: Schrodinger type equation}) with effective potentials equal to \cite{Kodama:2003kk}
\begin{equation} \label{eq: potential vector general}
V_V(r) = f(r)\left\{ \frac{k_V^2}{r^2} + \frac{(p^2-2p+4)K}{4r^2} - \frac{\lambda p(p-2)}{4}  + \frac{(p^2-4) M }{2 r^{p+1}}  \right\},
\end{equation} 
for vector type perturbations and 
\begin{equation} \label{eq: potential scalar general}
V_S(r) = f(r)\left\{ \frac{k_S^2 + \tfrac{p(p-2)K}{4} }{r^2} - \frac{\lambda (p^2-6p+8)}{4}  - \frac{(3p-2)(p-2) M }{2 r^{p+1}}  \right\},
\end{equation} 
for scalar type perturbations (see also Refs.\ \cite{Crispino:2000jx}). In Eqs.\ (\ref{eq: potential vector general}) and (\ref{eq: potential scalar general}), $k_V^2$ and $k_S^2$ are the eigenvalues of the vector type and scalar type tensor harmonics on the $p$-dimensional Einstein space with line element ${\rm d} \sigma^2_p$ and the function $f(r)$ is given by \cite{Kodama:2003kk}
\begin{equation}
 f(r) = K - \lambda r^2 - \frac{2 M}{r^{p-1}},
\end{equation} 
where $K= \pm 1,0$; $\lambda= 2 \Lambda / (p(p+1))$, $\Lambda $ is the cosmological constant, and the quantity $M$ is related to the mass of the spacetime with metric (\ref{e: metric general}).

\section{WKB approximation}
\label{Section WKB}

To calculate the QNMs of a spacetime there are several numerical and semi-analytical methods, (see Refs.\ \cite{Kokkotas:1999bd} for reviews of these methods). Here using the sixth order WKB method  \cite{Konoplya:2003ii}, \cite{Schutz-Will:1985}, \cite{Iyer:1986np}, we calculate the QN frequencies of the electromagnetic field propagating in $D$-dimensional Schwarzschild and SdS black holes in Sects.\ \ref{Section Schwarzschild} and \ref{Section SdS} respectively. Thus it is convenient to make a brief discussion of the WKB method. 

The WKB approximation is based on the analogy that exists between the scattering of waves on the peak of a potential barrier in Quantum Mechanics and the QNMs of a black hole. The use of the WKB approximation for calculating the QN frequencies of black holes was proposed by Schutz and Will in Ref.\ \cite{Schutz-Will:1985}, and their formula was generalized to third order beyond the eikonal approximation by Iyer and Will \cite{Iyer:1986np} and more recently it was extended to sixth order beyond the eikonal approximation by Konoplya in Ref.\ \cite{Konoplya:2003ii}. 

In the sixth order WKB approximation the QN frequencies of a black hole are given by the formula \cite{Konoplya:2003ii}, \cite{Iyer:1986np}
\begin{equation} \label{e: WKB frequencies}
\omega^2 = V_0 - i (-2V_0^{\prime \prime})^{1/2} (n + \tfrac{1}{2} + \Lambda_2 + \Lambda_3 + \Lambda_4 + \Lambda_5 + \Lambda_6),
\end{equation} 
where $n$ is the mode number, $V_0$ is the maximum value of the effective potential and $V_0^{\prime \prime}$ the value of its second derivative at the maximum. The expressions for $\Lambda_2$ and $\Lambda_3$ are given in Eqs.\ (15.a) and (15.b) of Ref.\ \cite{Iyer:1986np} and those for $\Lambda_4$, $\Lambda_5$, and $\Lambda_6$ appear in Appendix A of Ref.\ \cite{Konoplya:2003ii}.

It is well known that the third order WKB approximation yields results with an accuracy of $1\%$ for both real and imaginary parts of the QN frequencies as $n < l$, where $l$ is the angular momentum number (multipole number). Previous results showed that using the sixth order WKB we can get more accurate values for the QN frequencies and sometimes the findings are identical to those obtained by means of more complex numerical methods \cite{Konoplya:2003ii}. Therefore we believe that the WKB approximation is an efficient method to calculate the low mode number QNMs of several black holes and it has been used in several references, for example \cite{Konoplya:2003ii}, \cite{Cho:2007zi}, \cite{Kanti:2005xa}, \cite{Konoplya:2003dd}, \cite{Iyer:1986np}, \cite{Iyer:1986nq}, \cite{Kokkotas:1988fm}, \cite{Cho:2003qe}--\cite{Otsuki:2004yo}.

It was noted that if the number of dimensions increases then the WKB approximation converges more slowly \cite{Konoplya:2003ii}, \cite{Konoplya:2003dd}. Moreover, if the number of dimensions increases then the convergence of the WKB method for the modes with $n=l$ or $n=l-1$ is slower and in some cases we cannot assert that the WKB formula (\ref{e: WKB frequencies}) gives reliable results for the values of the QN frequencies of some fields.

\section{$D$-dimensional Schwarzschild black hole}
\label{Section Schwarzschild}

Using the sixth order WKB approximation \cite{Konoplya:2003ii}, \cite{Iyer:1986np}, in this section we compute the QN frequencies of the electromagnetic perturbation propagating in $D$-dimensional Schwarzschild black hole for $D=5,6,7,8,9,10$. Noticing that for these spacetimes $\lambda=0$, $k_S^2=k_V^2+1=l(l+p-1)$, the effective potentials for the vector type (\ref{eq: potential vector general}) and scalar type (\ref{eq: potential scalar general}) electromagnetic fields take the form
\begin{align} \label{e: Schwarzschild vector}
V_V(r) &= f(r)\left\{ \frac{l(l+p-1) + \tfrac{p(p-2)}{4} }{r^2}  + \frac{(p-2)(p+2) M }{2 r^{p+1}}  \right\},  \\
V_S(r) &= f(r)\left\{ \frac{l(l+p-1) + \tfrac{p(p-2)}{4} }{r^2} - \frac{ (3p-2)(p-2) M }{2 r^{p+1}}  \right\}, \label{e: Schwarzschild scalar}
\end{align} 
where for $(p+2)$-dimensional Schwarzschild background the function $f(r)$ of formula (\ref{e: metric general}) is equal to
\begin{equation}
f(r) = 1 -\frac{2 M}{r^{p-1}},
\end{equation} 
and in this spacetime the quantity ${\rm d} \sigma_p^2$, which also appear in formula (\ref{e: metric general}), is the line element of a $p$-dimensional sphere.

In the present section we take the constant $M$ equal to $1$ ($M=1$). An analysis of the plots for the effective potentials (\ref{e: Schwarzschild vector}) and (\ref{e: Schwarzschild scalar}) shows that these take the form of  potential barriers such that at the horizon and at infinity these tend to constant values.

For the cases that we study below, from the graphs of the effective potentials for the vector type perturbations we note that these are positive definite for $r>r_H$, where $r_H$ is the radius of the black hole horizon. 

In the cases that we analyze below for the scalar type perturbations the effective potentials are positive definite for $r>r_H$ except when the angular momentum number takes the value $l=1$ and the spacetime dimension is $D=7,8,9,10$, and when the angular momentum number is $l=2$ and the spacetime dimension is $D=10$. 

Taking into account these facts and using the formula (\ref{e: WKB frequencies}), which determines the QN frequencies in the sixth order WKB approximation, we find for the vector type and scalar type electromagnetic fields the results presented in Tables \ref{tab: Schwarzschild p=3}--\ref{tab: Schwarzschild p=8} when the spacetime is the $(p+2)$-dimensional Schwarzschild black hole with $p=3,4,5,6,7,8$. In these tables $\omega^s_R$ and $\omega^s_I$ ($\omega^v_R$ and $\omega^v_I$) denote the real and imaginary parts of the QN frequencies for scalar (vector) type electromagnetic perturbations.

For the cases in which the effective potentials have in addition to the usual maximum, a minimum outside the horizon of the black hole, we expect that the WKB approximation is not valid \cite{Konoplya:2003ii}, \cite{Iyer:1986np}; but even in these cases, using formula (\ref{e: WKB frequencies}) we calculate the QN frequencies since our numerical results for the third, fourth, fifth, and sixth order WKB approximation converge well and we give these values in Tables \ref{tab: Schwarzschild p=5}--\ref{tab: Schwarzschild p=8}. We only distinguish with an asterisk $\,^{*}\,$ the QN frequencies of the scalar type electromagnetic perturbations that we get for these values of the multipole number and spacetime dimension.  

In $D$-dimensional Schwarzschild black hole for the QN frequencies of the scalar type gravitational perturbations a similar method was used by Berti, et al.\ in Ref.\ \cite{Berti:2003si} for the cases in which the effective potentials for the gravitational perturbations of scalar type are not positive definite.

As far as we know there are few results in the literature on the low mode number QN frequencies of the electromagnetic field propagating in $D$-dimensional Schwarzschild background. We know those presented in Fig.\ 3 and Table V of Ref.\ \cite{Cardoso:2003vt} for the vector type perturbations propagating in a five-dimensional Schwarzschild spacetime. For low mode numbers our values are in good agreement with those of Cardoso, et al.\ \cite{Cardoso:2003vt} (see our Table \ref{tab: Schwarzschild p=3}). In the present work and the previously mentioned reference the mass is measured in different units. The relation between both sets of values for the QN frequencies is
\begin{equation}
\omega^v = \frac{\omega_{C}}{\sqrt{2} \, 2 \pi},
\end{equation}  
where $\omega^v$ are the values of the QN frequencies that we calculate here and $\omega_{C}$ are the values provided by Cardoso et al.\ in Ref.\ \cite{Cardoso:2003vt}.

For other recently published QN frequencies of vector type electromagnetic field propagating in $D$-dimensional Schwarzschild black hole for $D=5,6,7,$ $8,9,10$, $l=2$, and $n=0$ see the first row in Tables III and V of Ref.\ \cite{Konoplya:2007jv}.\footnote{In Ref.\ \cite{Konoplya:2007jv} Konoplya and Zhidenko call to the perturbations whose QN frequencies appear in their Table V, ``gravitational perturbations of vector (``+'') type''. In $D$-dimensional Schwarzschild and SdS black holes they correspond to electromagnetic fields of vector type \cite{Kodama:2003kk}.} Further, we note that in the previous references the QN frequencies for the electromagnetic perturbations of scalar type were not calculated.

From Tables \ref{tab: Schwarzschild p=3}--\ref{tab: Schwarzschild p=8}, for the values of the QN frequencies for the vector type and scalar type electromagnetic perturbations we find that the real and imaginary parts of these frequencies satisfy\footnote{In this paper the imaginary parts of the QN frequencies are negative numbers and except for the QN frequencies (\ref{e: topological QN vector}), (\ref{e: topological QN scalar}), and (\ref{e: quasinormal frequency vector}) we only consider the frequencies with positive real parts.}
\begin{align} \label{e: inequalities QN}
 |\omega^s_R| < |\omega^v_R|, \qquad \qquad 
|\omega^s_I| < |\omega^v_I| .
\end{align}
Moreover, from our results for the imaginary parts we notice that the inequality (\ref{e: inequalities QN}) is not valid only for $p=4$, $l=n=1$ (see Table \ref{tab: Schwarzschild p=4}), but it is probable that the numerical errors are the causes of the atypical behavior that we find for these values of $p$, $l$, and $n$, since for $n=l$ the WKB method converges more slowly, as we mentioned before.

Thus the inequalities (\ref{e: inequalities QN}) are similar to those that satisfy the low mode number QNMs of the gravitational perturbations propagating in $D$-dimen\-sion\-al Schwarzschild background \cite{Berti:2003si}, \cite{Konoplya:2003dd}. Hence in similar way to the gravitational perturbations, we find that the vector type electromagnetic fields have a greater oscillation frequency and larger damping rate than the scalar type electromagnetic perturbations (except for a case).

From Tables \ref{tab: Schwarzschild p=3}--\ref{tab: Schwarzschild p=8} we infer that for fixed $p$, $l$, and for both types of electromagnetic perturbations the real and imaginary parts of the QN frequencies decrease as the mode number increases. 

In Sect.\ \ref{Section WKB} we commented that in some previous works was observed that as the number of dimensions increases and $n=l$ or $n=l-1$, the WKB approximation converges more slowly. In our computations we have obtained that for $D=8,9,10$, the numerical values of the QN frequencies do not converge for some modes with $n=l$ or $n=l-1$. For these cases the corresponding cell of the table is left blank (for example see Tables \ref{tab: Schwarzschild p=6}--\ref{tab: Schwarzschild p=8}). In these tables we also note that for higher values of $p$ the number of cells left blank in the corresponding table increases. Thus as $p$ increases the WKB formula (\ref{e: WKB frequencies}) converges more slowly. Hence our results suppport the claims of previous works.
 
Comparing our values with those of Tables I and II of Ref.\ \cite{Konoplya:2003dd} we find that the real $\omega_{R,G}^v$ and imaginary $\omega_{I,G}^v$ parts of the QN frequencies for vector type gravitational perturbations satisfy\footnote{We recall that $\omega^v_I$, $\omega^s_I$, $\omega_{I,G}^v$, and $\omega_{I,G}^s$ are negative quantities.}
\begin{equation} \label{e: inequality vector}
\omega^v_R > \omega_{R,G}^v, \qquad \qquad \omega^v_I <  \omega_{I,G}^v,
\end{equation} 
for $n=0$, $p=3,4,5,6,7,8$ and $l=2$, or $l=3$. Furthermore, the real $\omega_{R,G}^s$ and imaginary $\omega_{I,G}^s$ parts of the QN frequencies for scalar type gravitational perturbations satisfy
\begin{equation} \label{e: inequality scalar}
\omega^s_R > \omega_{R,G}^s, \qquad \qquad \omega^s_I <  \omega_{I,G}^s,
\end{equation} 
for $l=3$, $n=0$, and $p=3,4,5,6,7$. Thus for both types of perturbations and for these values of $p$, $n$, and $l$, the oscillation frequency and the damping rate of the electromagnetic perturbations are larger than those for the gravitational perturbations of the same type. 

\begin{figure}[htbp]
\begin{center}
\caption{Dependence of the QN frequencies on the dimension of  Schwarzschild spacetime for the \textbf{vector type} electromagnetic perturbations with $n=0$ and $l=2$ (crosses), $l=3$ (squares), and $l=4$ (circles).}
\psfrag{wR}{$\omega_R^v$}
\psfrag{wI}{$\omega_I^v$}
\includegraphics[clip,scale=.44]{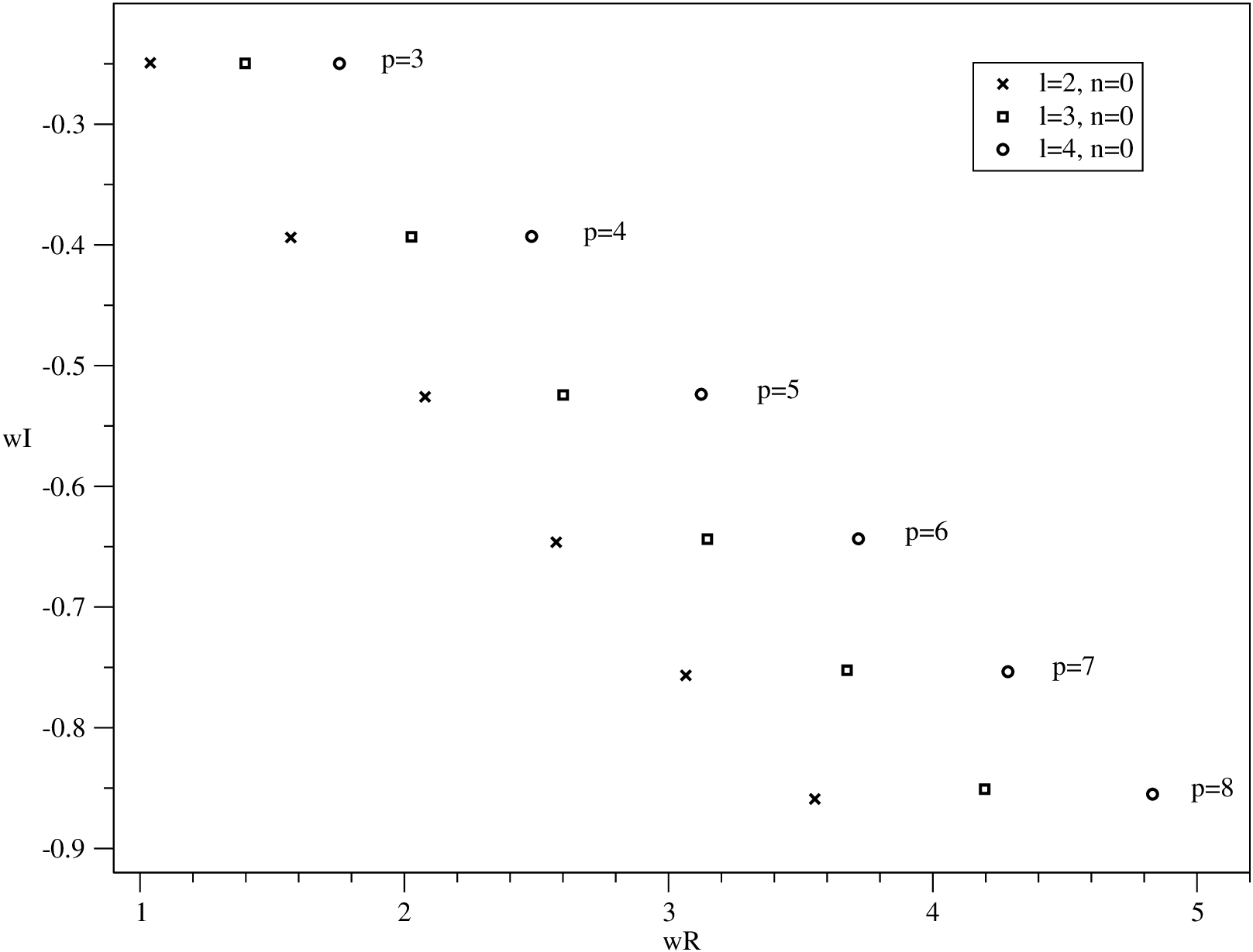}
\label{fig: s vector dimension n=0}
\end{center}
\end{figure}

\begin{figure}[htbp]
\begin{center}
\caption{Dependence of the QN frequencies on the dimension of Schwarzschild spacetime for the \textbf{vector type} electromagnetic perturbations with $n=1$ and $l=2$ (crosses), $l=3$ (squares), and $l=4$ (circles).}
\psfrag{wI}{$\omega_I^v$}
\psfrag{wR}{$\omega_R^v$}
\includegraphics[clip,scale=.44]{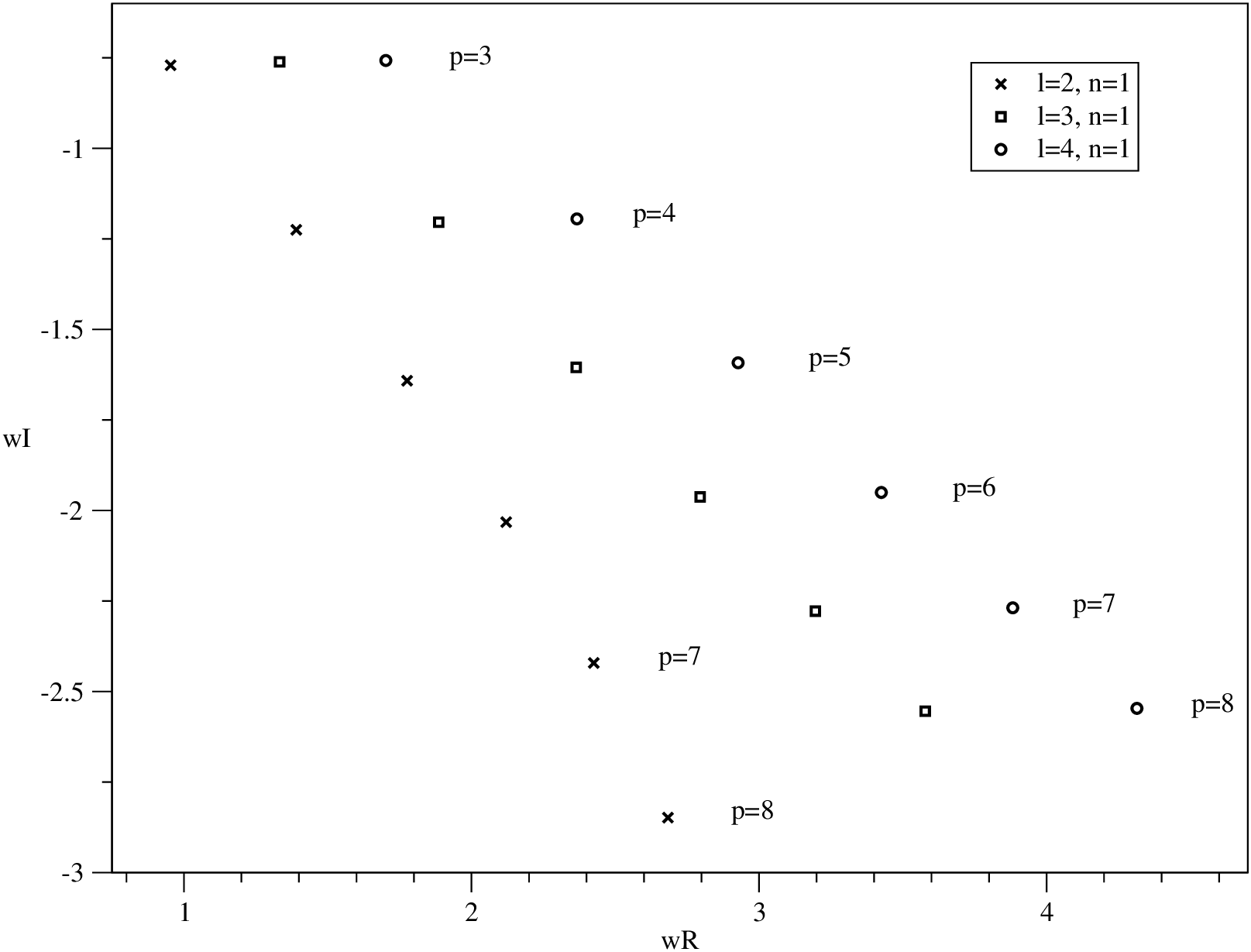}
\label{fig: s vector dimension n=1}
\end{center}
\end{figure}

\begin{figure}[htbp]
\begin{center}
\caption{Dependence of the QN frequencies on the dimension of Schwarzschild spacetime for the \textbf{scalar type} electromagnetic perturbations with $n=0$ and $l=2$ (crosses), $l=3$ (squares), and $l=4$ (circles).}
\psfrag{wR}{$\omega_R^s$}
\psfrag{wI}{$\omega_I^s$}
\includegraphics[clip,scale=.43]{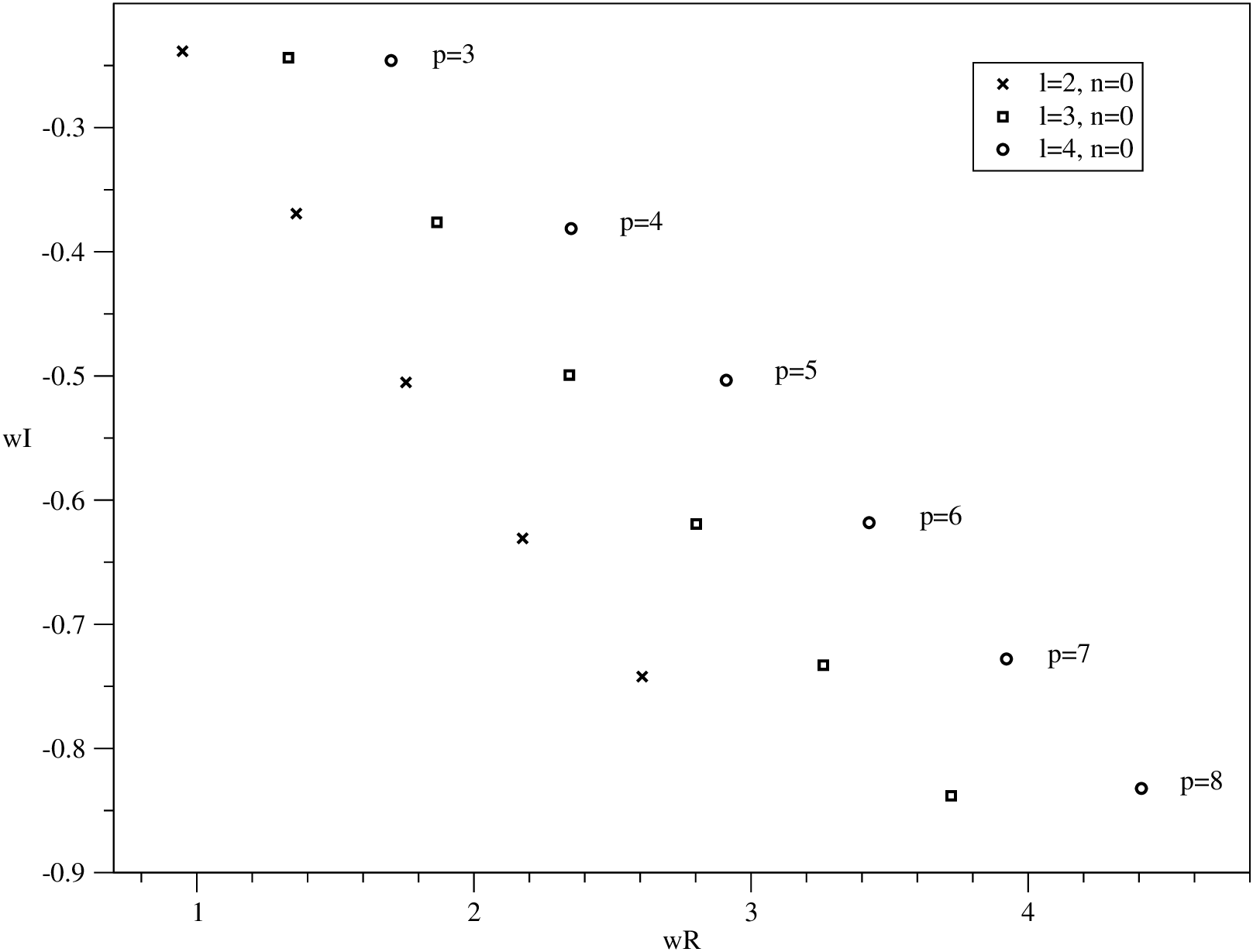}
\label{fig: scalar dimension n=0}
\end{center}
\end{figure}

\begin{figure}[htbp]
\begin{center}
\caption{Dependence of the QN frequencies on the dimension of Schwarzschild spacetime for the \textbf{scalar type} electromagnetic perturbations with $n=1$ and $l=2$ (crosses), $l=3$ (squares), and $l=4$ (circles).}
\psfrag{wI}{$\omega_I^s$}
\psfrag{wR}{$\omega_R^s$}
\includegraphics[clip,scale=.43]{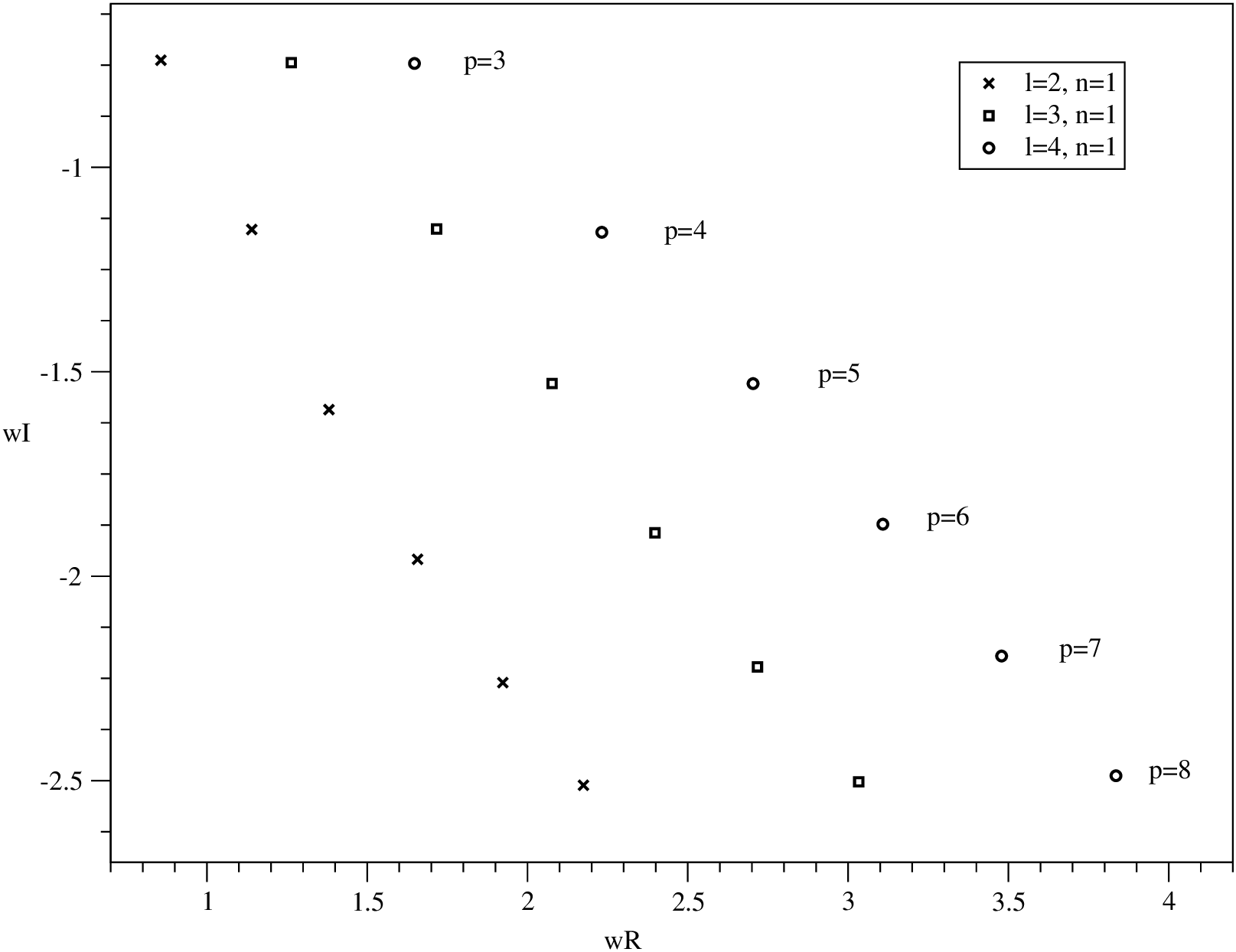}
\label{fig: scalar dimension n=1}
\end{center}
\end{figure}

For fixed angular momentum and mode numbers, as the dimension of Schwarzschild background increases from $D=5$ up to $D=10$ the real and imaginary parts of the QN frequencies for vector type and scalar type electromagnetic perturbations are plotted in Figs.\ \ref{fig: s vector dimension n=0}--\ref{fig: scalar dimension n=1}. These figures show that for fixed $p$ and $n$ and for both types of electromagnetic perturbations the real parts of the QN frequencies are different for different values of the angular momentum number (the real part increases as the angular momentum number increases). The reason of this behavior is provided in Subsect.\ \ref{Subsection Large l} below. Also, for fixed $l$ and $n$ the real parts of the QN frequencies increase as the number of dimensions increases.

Moreover for the cases studied here and for fixed $p$ and $n$, Figs.\ \ref{fig: s vector dimension n=0}--\ref{fig: scalar dimension n=1} show that the imaginary parts of the QN frequencies are almost equal (in the scale of these figures) for different values of the angular momentum number, except for the imaginary parts of the QN frequencies for the vector type perturbations with parameters $l=2$, $n=1$, which are manifestly different from those with $l=3$, or $l=4$, and $n=1$ in $D=9$ and $D=10$ dimensions (see Fig.\ \ref{fig: s vector dimension n=1}). Furthermore for fixed $l$ and $n$ the imaginary parts of the QN frequencies decrease as the number of dimensions increases.

Finally, we do not compare our results for the electromagnetic field with those of Table III in Ref.\ \cite{Kanti:2005xa}. In the previous reference are studied the QNMs of the electromagnetic field propagating in a four-dimensional background and the extra spatial dimensions only change the effective metric of the four-dimensional spacetime.

\section{$D$-dimensional Schwarzschild de Sitter black hole}
\label{Section SdS}

In Refs.\ \cite{Kanti:2005xa}, \cite{Konoplya:2003dd}, \cite{Zhidenko:2003wq}--\cite{Otsuki:2004yo} the WKB method has been used previously to compute the QN frequencies of different fields moving in four and $D$-dimensional SdS black holes. As in the previous section for Schwarzschild background, here we use the sixth order WKB approximation of Ref.\ \cite{Konoplya:2003ii} to compute the QN frequencies of the electromagnetic field propagating in $D$-dimensional SdS black hole for $D=5,6,7,8$, filling a gap in the literature.

For $D$-dimensional SdS spacetime the effective potentials (\ref{eq: potential vector general}) and (\ref{eq: potential scalar general}) of the electromagnetic field take the form ($\lambda \neq 0$)
\begin{align} \label{e: SdS vector}
V_V(r) & = f(r)\left\{ \frac{l(l+p-1) + \tfrac{p(p-2)}{4} }{r^2} - \frac{\lambda p(p-2)}{4}  + \frac{(p^2-4) M }{2 r^{p+1}}  \right\}, \\
\label{e: SdS scalar}
V_S(r) & = f(r)\left\{ \frac{  l(l+p-1) + \tfrac{p(p-2)}{4} }{r^2} - \frac{ \lambda (p^2-6p+8)}{4}  - \frac{(3p^2-8p+4) M }{2 r^{p+1}}  \right\},
\end{align} 
where the function $f(r)$ of formula (\ref{e: metric general}) for this spacetime is equal to
\begin{equation} \label{e: function SdS}
f(r) = 1 -\frac{2 M}{r^{p-1}} - \lambda r^2,
\end{equation} 
and ${\rm d} \sigma_p^2$ denotes the line element of a $p$-dimen\-sion\-al sphere as for Schwarzschild black hole.

Taking $M=1$ in formula (\ref{e: function SdS}), we find that the metric of formula (\ref{e: metric general}) describes a black hole with a cosmological horizon if the parameter $\lambda$ satisfies the following condition\footnote{The parameter $\lambda$ is related to the cosmological constant.} \cite{Kodama:2003kk}
\begin{equation} \label{e: SdS black hole condition}
 \frac{p-1}{p+1} \frac{1}{(p+1)^{2/(p-1)}}  > \lambda .
\end{equation} 
We assume that the $D$-dimensional SdS black holes that we study below in this section satisfy this condition. 

In Ref.\ \cite{Konoplya:2007jv} was studied numerically the linear stability of $D$-dimen\-sional SdS black holes against gravitational perturbations of scalar type, since for this type of gravitational perturbations the effective potentials are not positive definite. No gravitational instability was found in the previous reference.

For ($p+2$)-dimensional SdS spacetime with $p=3,4,5,6,$ the electromagnetic field with $l=2,3,4,5,$ and the values of $\lambda$ that we study here, the graphs of the effective potentials (\ref{e: SdS vector}) and (\ref{e: SdS scalar}), show that they are definite positive as the radial coordinate varies between the cosmological and black hole horizons. Also, the same graphs show that these effective potentials tend to constant values at the cosmological and black hole horizons.

\begin{figure}[htbp]
\begin{center}
\caption{Real parts of the QN frequencies for the \textbf{vector type} electromagnetic perturbations moving in SdS black hole for $l=3$, $n=0$, $p=3$ (crosses), $p=4$ (squares), $p=5$ (circles), and $p=6$ (diamonds).}
\psfrag{lambda}{$\lambda$}
\psfrag{wR}{$\omega_R^v$}
\includegraphics[clip,scale=.43]{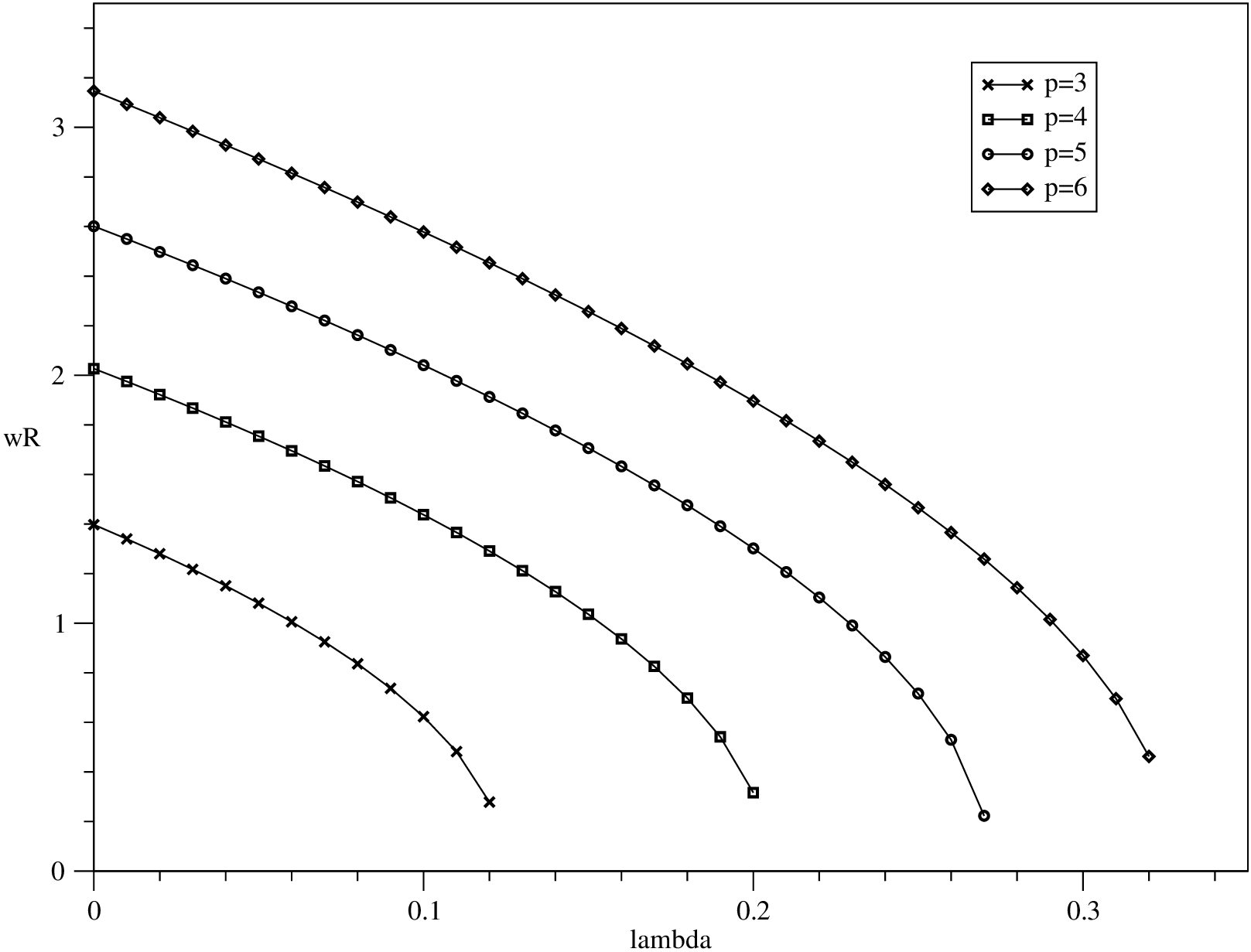}
\label{fig: Re-sds-vector-l=3}
\end{center}
\end{figure}

\begin{figure}[htbp]
\begin{center}
\caption{Imaginary parts of the QN frequencies for the \textbf{vector type} electromagnetic perturbations moving in SdS black hole for $l=3$, $n=0$, $p=3$ (crosses), $p=4$ (squares), $p=5$ (circles), and $p=6$ (diamonds).}
\psfrag{lambda}{$\lambda$}
\psfrag{wI}{$\omega_I^v$}
\includegraphics[clip,scale=.43]{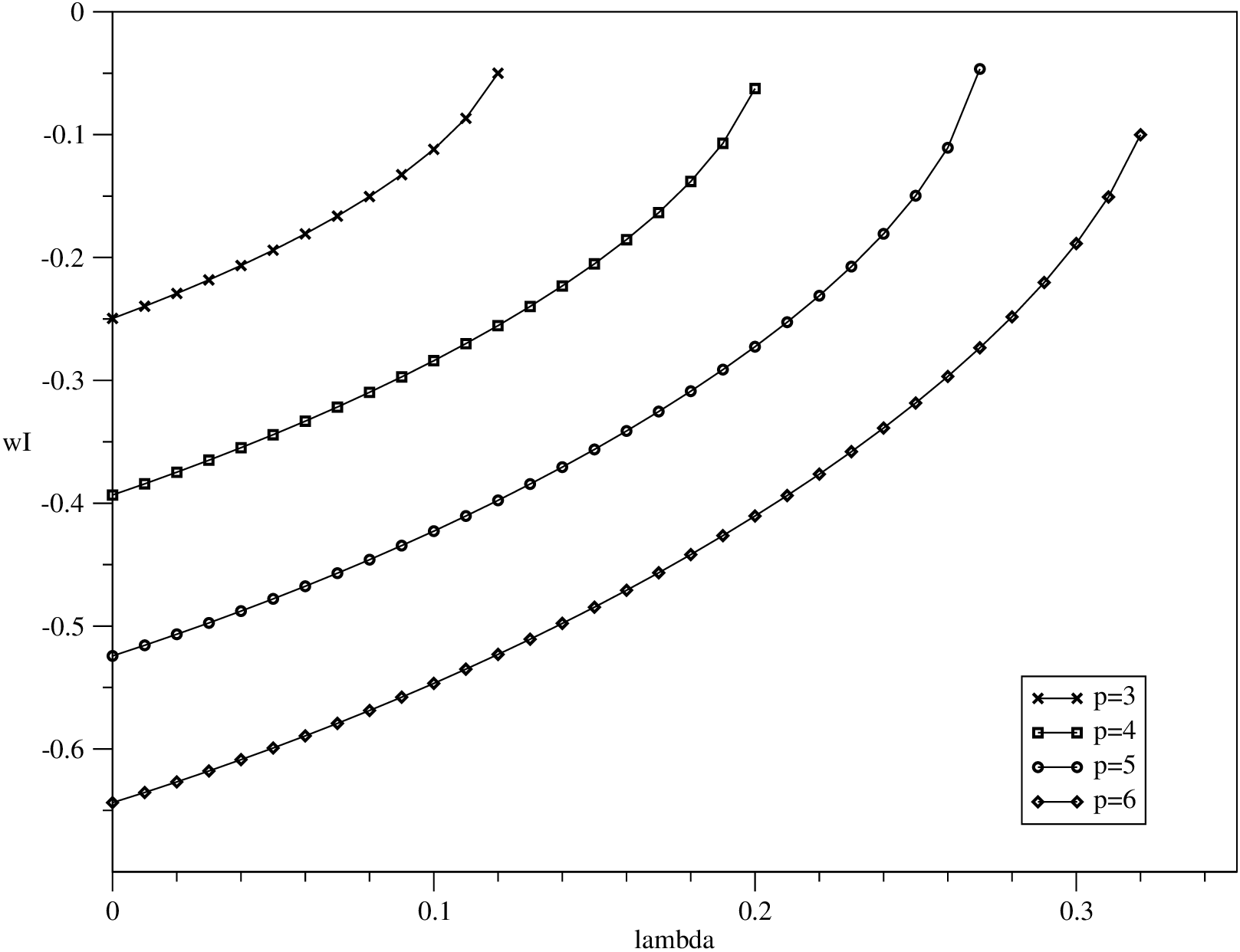}
\label{fig: Im-sds-vector-l=3}
\end{center}
\end{figure}

\begin{figure}[htbp]
\begin{center}
\caption{Real parts of the QN frequencies for the \textbf{vector type} electromagnetic perturbations moving in SdS black hole for $l=4$, $n=0$, $p=3$ (crosses), $p=4$ (squares), $p=5$ (circles), and $p=6$ (diamonds).}
\psfrag{lambda}{$\lambda$}
\psfrag{wR}{$\omega_R^v$}
\includegraphics[clip,scale=.43]{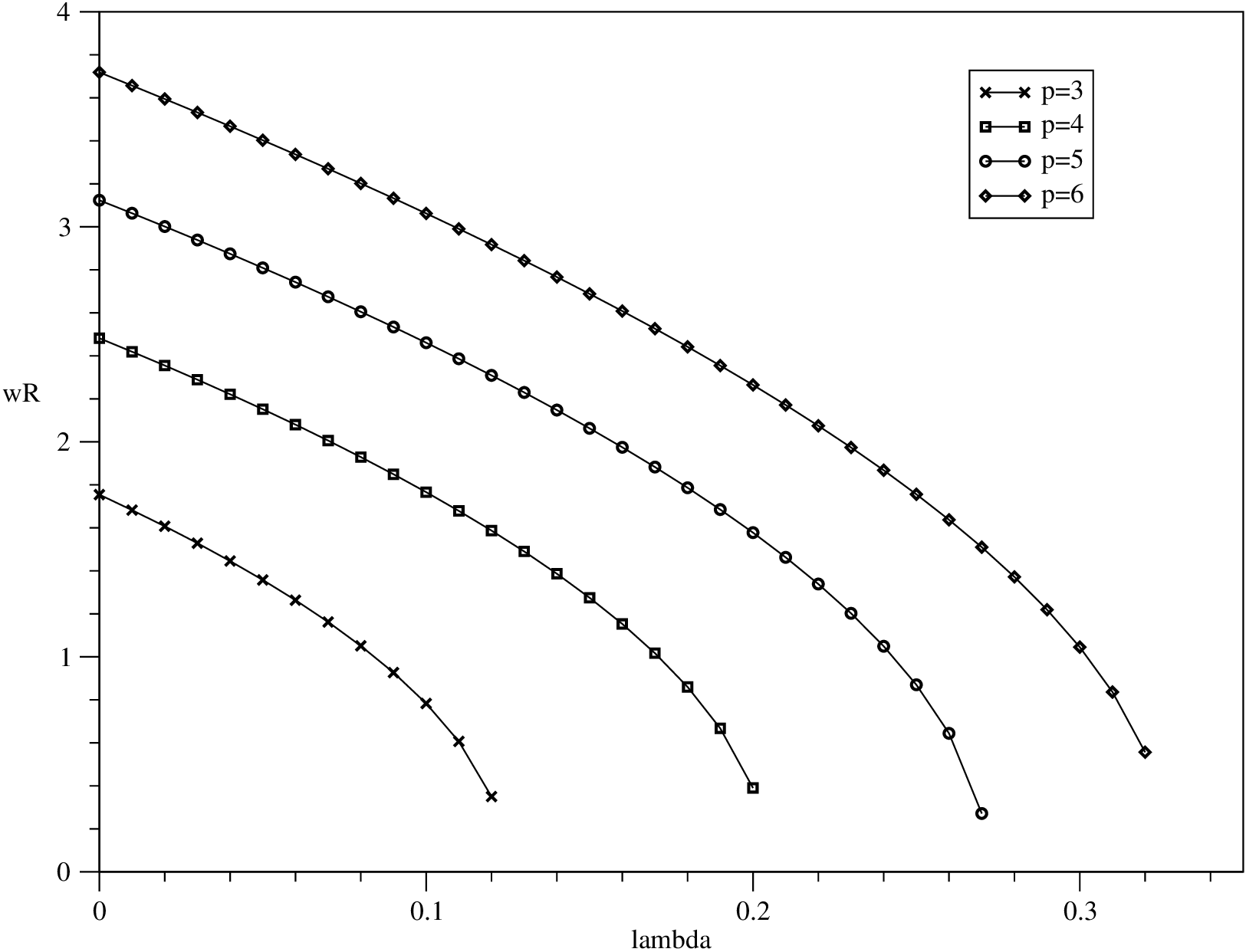}
\label{fig: Re-sds-vector-l=4}
\end{center}
\end{figure}

\begin{figure}[htbp]
\begin{center}
\caption{Imaginary parts of the QN frequencies for the \textbf{vector type} electromagnetic perturbations moving in SdS black hole for $l=4$, $n=0$, $p=3$ (crosses), $p=4$ (squares), $p=5$ (circles), and $p=6$ (diamonds).}
\psfrag{lambda}{$\lambda$}
\psfrag{wI}{$\omega_I^v$}
\includegraphics[clip,scale=.43]{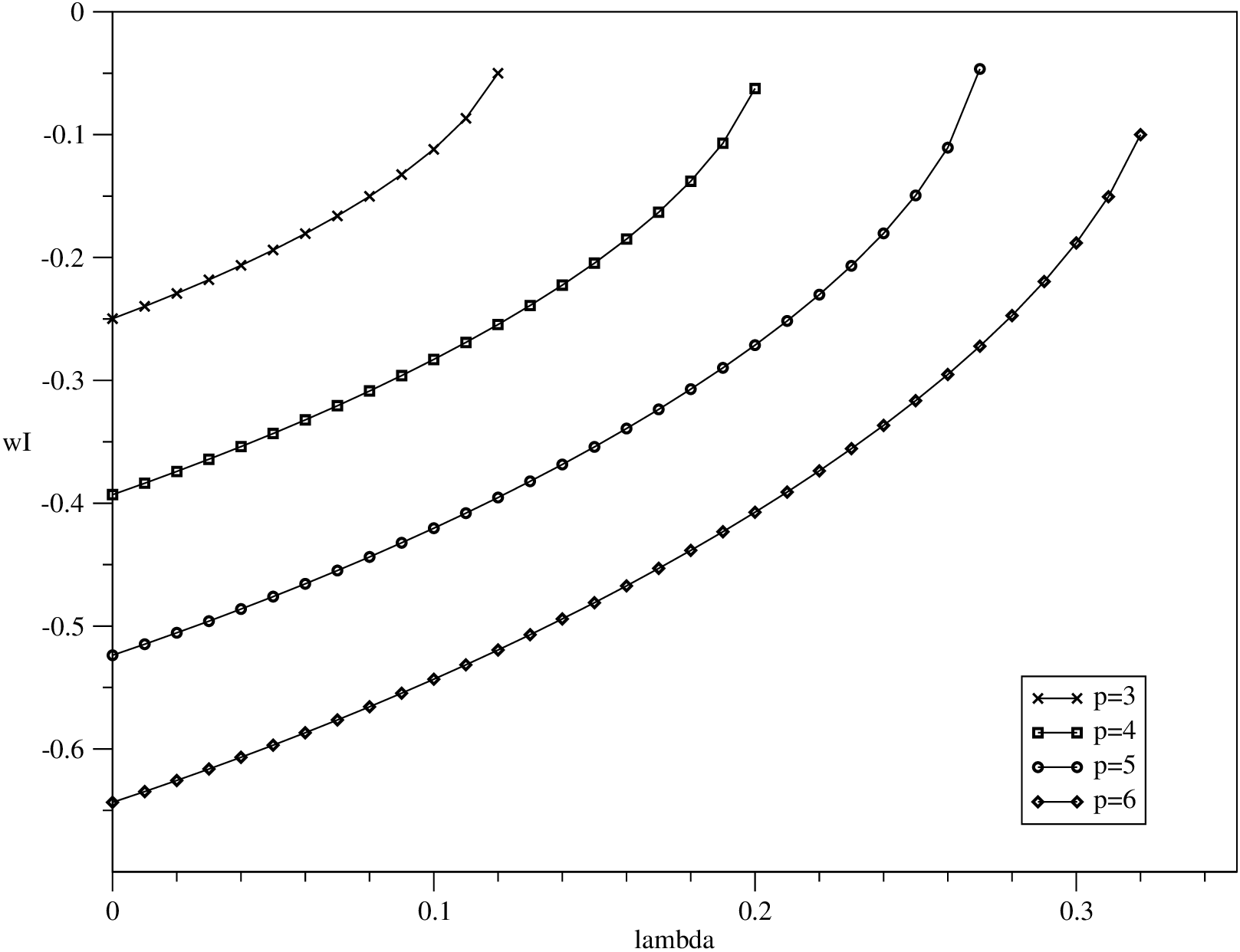}
\label{fig: Im-sds-vector-l=4}
\end{center}
\end{figure}

\begin{figure}[htbp]
\begin{center}
\caption{Real parts of the QN frequencies for the \textbf{scalar type} electromagnetic perturbations moving in SdS black hole for $l=3$, $n=0$, $p=3$ (crosses), $p=4$ (squares), $p=5$ (circles), and $p=6$ (diamonds).}
\psfrag{lambda}{$\lambda$}
\psfrag{wR}{$\omega_R^s$}
\includegraphics[clip,scale=.43]{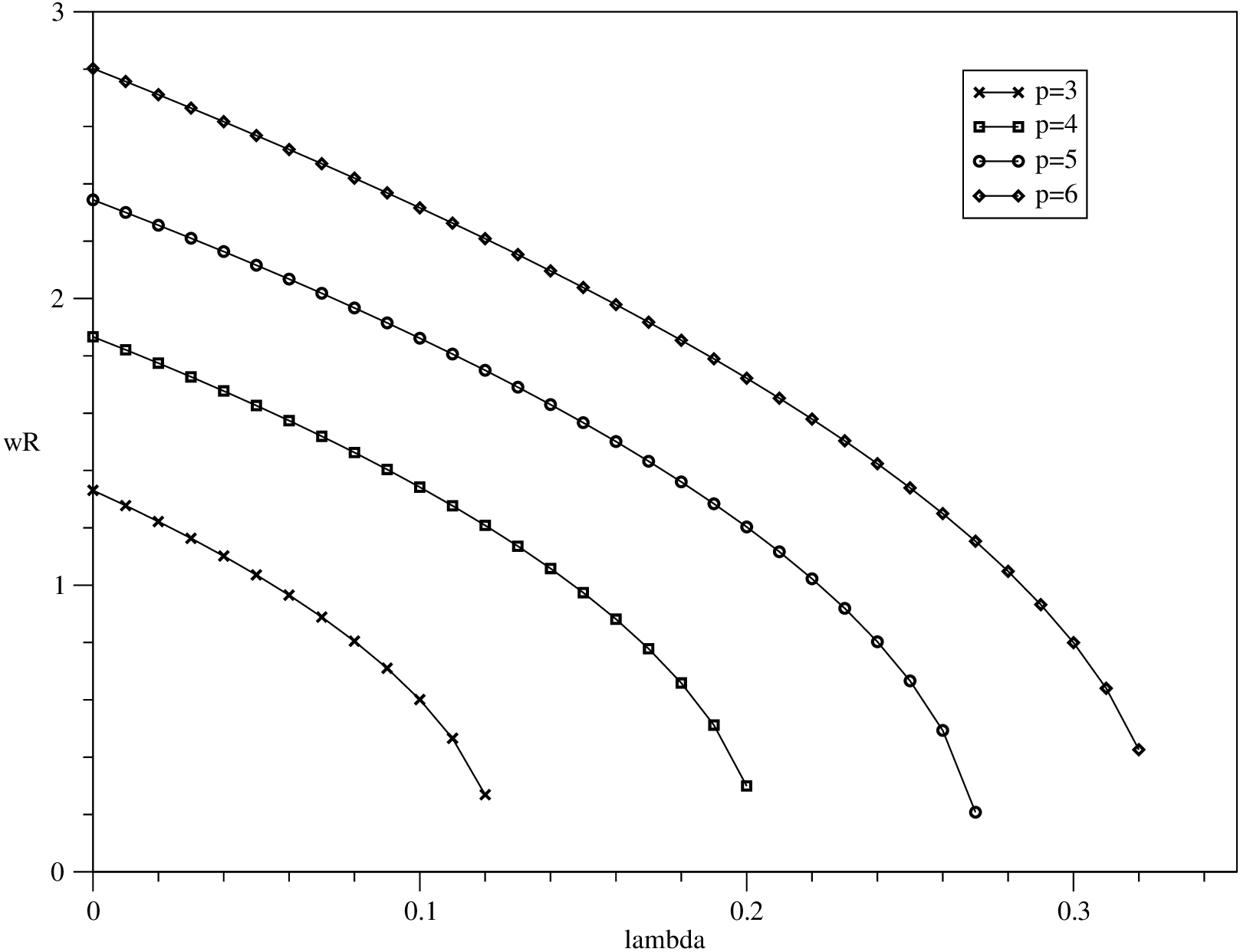}
\label{fig: Re-sds-scalar-l=3}
\end{center}
\end{figure}

\begin{figure}[htbp]
\begin{center}
\caption{Imaginary parts of the QN frequencies for the \textbf{scalar type} electromagnetic perturbations moving in SdS black hole for $l=3$, $n=0$, $p=3$ (crosses), $p=4$ (squares), $p=5$ (circles), and $p=6$ (diamonds).}
\psfrag{lambda}{$\lambda$}
\psfrag{wI}{$\omega_I^s$}
\includegraphics[clip,scale=.43]{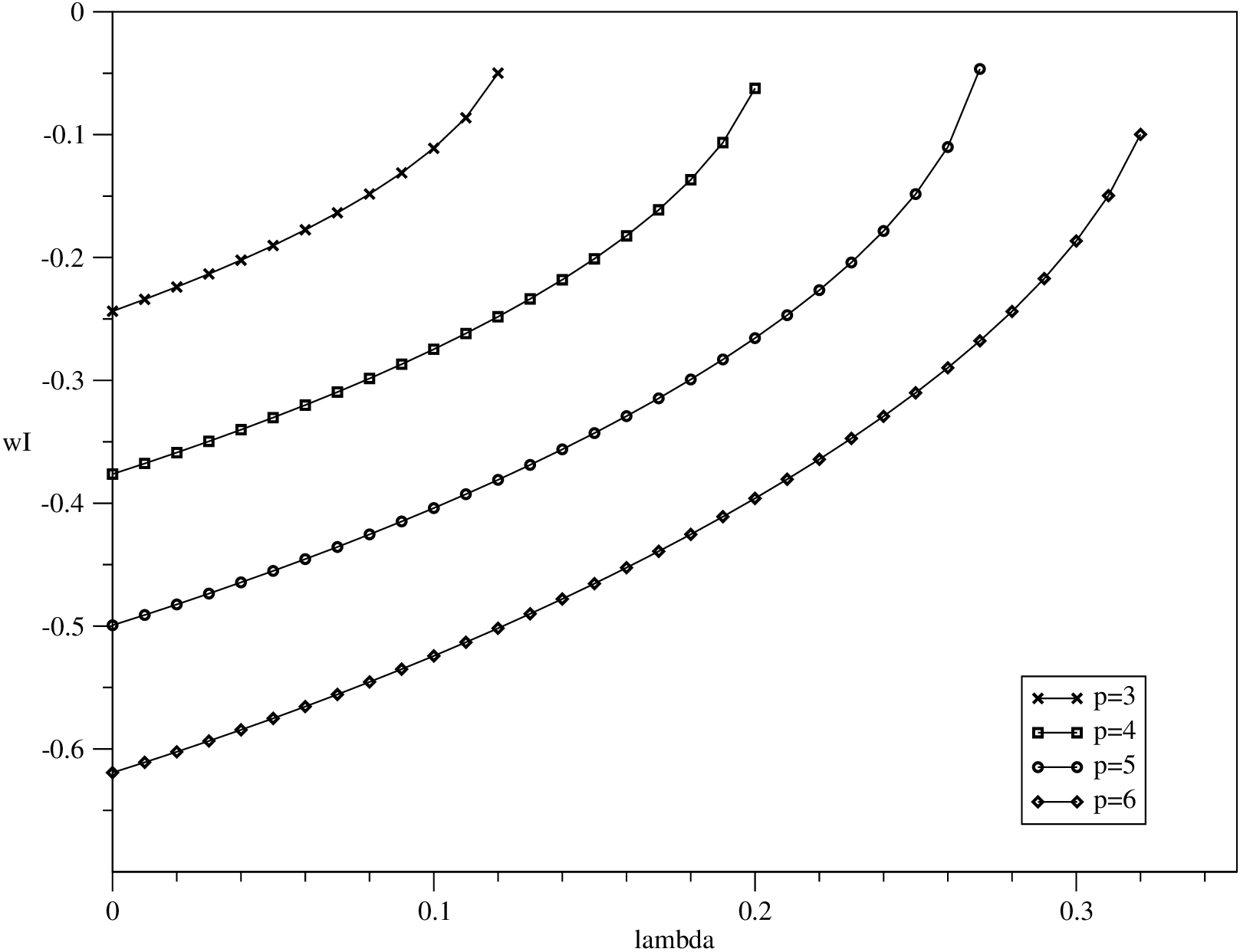}
\label{fig: Im-sds-scalar-l=3}
\end{center}
\end{figure}

\begin{figure}[htbp]
\begin{center}
\caption{Real parts of the QN frequencies for the \textbf{scalar type} electromagnetic perturbations moving in SdS black hole for $l=4$, $n=0$, $p=3$ (crosses), $p=4$ (squares), $p=5$ (circles), and $p=6$ (diamonds).}
\psfrag{lambda}{$\lambda$}
\psfrag{wR}{$\omega_R^s$}
\includegraphics[clip,scale=.43]{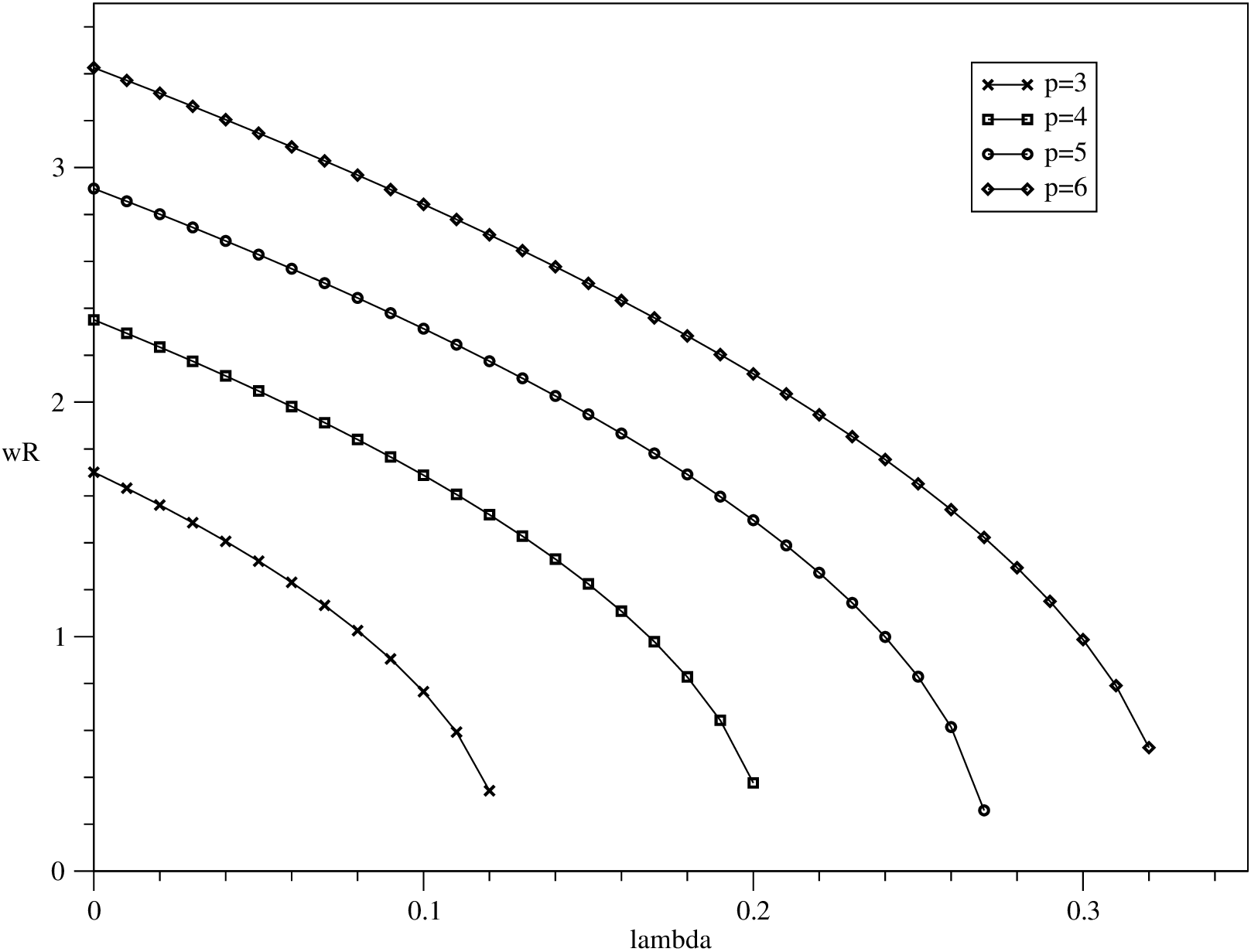}
\label{fig: Re-sds-scalar-l=4}
\end{center}
\end{figure}

\begin{figure}[htbp]
\begin{center}
\caption{Imaginary parts of the QN frequencies for the \textbf{scalar type} electromagnetic perturbations moving in SdS black hole for $l=4$, $n=0$, $p=3$ (crosses), $p=4$ (squares), $p=5$ (circles), and $p=6$ (diamonds).}
\psfrag{lambda}{$\lambda$}
\psfrag{wI}{$\omega_I^s$}
\includegraphics[clip,scale=.43]{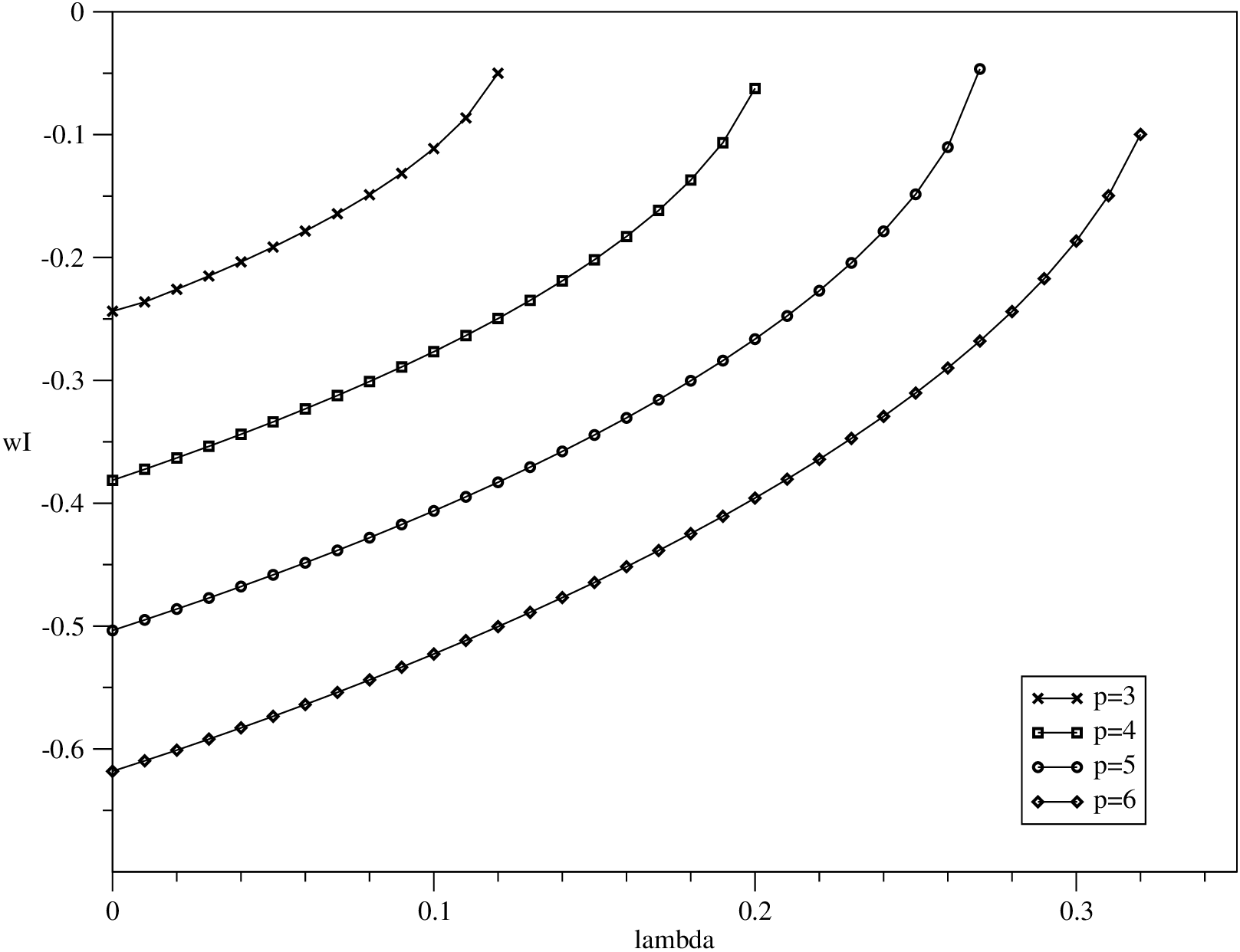}
\label{fig: Im-sds-scalar-l=4}
\end{center}
\end{figure}

From formula (\ref{e: WKB frequencies}) and the effective potential (\ref{e: SdS vector}) we get that the real and imaginary parts of the QN frequencies for the vector type electromagnetic perturbations depend on the value of the parameter $\lambda$ as plotted in Figs.\ \ref{fig: Re-sds-vector-l=3} and \ref{fig: Im-sds-vector-l=3} ($l=3$, $n=0$, and $p=3,4,5,6$) and Figs.\ \ref{fig: Re-sds-vector-l=4} and \ref{fig: Im-sds-vector-l=4} ($l=4$, $n=0$, and $p=3,4,5,6$). For the scalar type electromagnetic perturbations the real and imaginary parts of the QN frequencies depend on the parameter $\lambda$ as depicted in Figs.\ \ref{fig: Re-sds-scalar-l=3} and \ref{fig: Im-sds-scalar-l=3} ($l=3$, $n=0$, and $p=3,4,5,6$) and Figs.\ \ref{fig: Re-sds-scalar-l=4} and \ref{fig: Im-sds-scalar-l=4} ($l=4$, $n=0$, and $p=3,4,5,6$). 

Furthermore for both types of electromagnetic perturbations with $l=2$ and $l=5$ we also calculate the dependence on the parameter $\lambda$ of their QN frequencies for $n=0$ and $p=3,4,5,6$. The results that we found for the QN frequencies with these angular eigenvalues behave similarly to those depicted in Figs.\ \ref{fig: Re-sds-vector-l=3}--\ref{fig: Im-sds-scalar-l=4} for $l=3$ and $l=4$. For the same values of $p$ and $l$, but with $n=1$ we calculate the dependence on the parameter $\lambda$ of the QN frequencies for both types of electromagnetic perturbations. The obtained results behave similarly to those for $n=0$ that we have plotted in Figs.\ \ref{fig: Re-sds-vector-l=3}--\ref{fig: Im-sds-scalar-l=4}.

For $l=2,3,4,5$, $n=0,1$ and $p=3,4,5,6$, from the results depicted in Figs.\ \ref{fig: Re-sds-vector-l=3}--\ref{fig: Im-sds-scalar-l=4} and other that we do not plot here, we find that the absolute values of the real and imaginary parts of the QN frequencies for scalar and vector type perturbations satisfy the same inequalities (\ref{e: inequalities QN}) that the corresponding quantities for Schwarzschild black hole. Thus the oscillation frequency of the vector type electromagnetic perturbations is greater than that of the scalar type electromagnetic fields, similar to the gravitational perturbations propagating in $D$-dimensional SdS black hole \cite{Konoplya:2003dd}. Further, for these values of $p$, $l$, and $n$ the damping rate for vector type electromagnetic fields is greater than for scalar type electromagnetic fields, similar to the gravitational perturbations \cite{Konoplya:2003dd}.
 
Also from Figs.\ \ref{fig: Re-sds-vector-l=3}--\ref{fig: Im-sds-scalar-l=4} we note that for fixed $l$ and $n$ as the value of the parameter $\lambda$ increases, in some cases the absolute values of the real and imaginary parts of the QN frequencies decreases up to $80\%$. Thus the electromagnetic field damp more slowly and its period of oscillation increases as the quantity $\lambda$ grows from zero to its extremal value (see formula (\ref{e: SdS black hole condition})). This behavior is similar to that previously observed for fields of different spin moving in a four-dimensional SdS black hole \cite{Zhidenko:2003wq}, for gravitational perturbations in higher dimensional SdS spacetime \cite{Konoplya:2007jv}, \cite{Konoplya:2003dd}, and for massless fields propagating in Brane World SdS black hole \cite{Kanti:2005xa}.

Above we have calculated the QN frequencies of the electromagnetic field with $l=2,3,4,5$, propagating in $D$-dimensional SdS black hole for several values of $\lambda$. Observe that we have not provided the results of the QN frequencies for both types of electromagnetic perturbations with $l=1$. For this value of the angular momentum number and for some values of the parameter $\lambda$ the effective potentials of the scalar type electromagnetic field in $D=7$ and $D=8$ spacetime dimensions are not definite positive outside the black hole horizon (for this value of $l$, the effective potentials for vector type perturbations are definite positive). As in the previous section, for the electromagnetic field with $l=1$ propagating in $D$-dimensional SdS black hole, we compute, using again the WKB method, its low mode number QN frequencies. 

Our results for the vector type electromagnetic field show that the dependence upon $\lambda$ of the real and imaginary parts of its QN frequencies is similar to those depicted in Figs.\ \ref{fig: Re-sds-vector-l=3}--\ref{fig: Im-sds-scalar-l=4}. For scalar type electromagnetic perturbations the numerical values that we find for the fundamental mode $n=0$, show that the dependence on $\lambda$ of the QN frequencies is similar to the previous cases, but for $n=1$ the real and imaginary parts of the QN frequencies have a dependence on $\lambda$ different from the cases plotted in Figs.\ \ref{fig: Re-sds-vector-l=3}--\ref{fig: Im-sds-scalar-l=4}. The difference is more significant in the plot for the real parts of the QN frequencies.

Another curious fact about these QN frequencies with $l=1$ is that for some values of $\lambda$ the imaginary parts of those corresponding to electromagnetic fields of vector type can be larger than those of the scalar type electromagnetic perturbations. Thus for these values of the cosmological constant and multipole number, the scalar type electromagnetic perturbations are more damped than those of vector type, which is different from the behavior found in Schwarzschild black hole. Since the WKB method applies to definite positive potentials and the effective potential of scalar perturbations can be negative for $l=1$ outside the black hole horizon, we believe that these results deserve further investigation.

\begin{figure}[htbp]
\begin{center}
\caption{Dependence on the dimension of the QN frequencies for the \textbf{vector type} electromagnetic perturbations moving in SdS black hole with $\lambda=.070$, for $n=0$ and $l=2$ (crosses), $l=3$ (squares), and $l=4$ (circles).}
\psfrag{wR}{$\omega_R^v$}
\psfrag{wI}{$\omega_I^v$}
\includegraphics[clip,scale=.43]{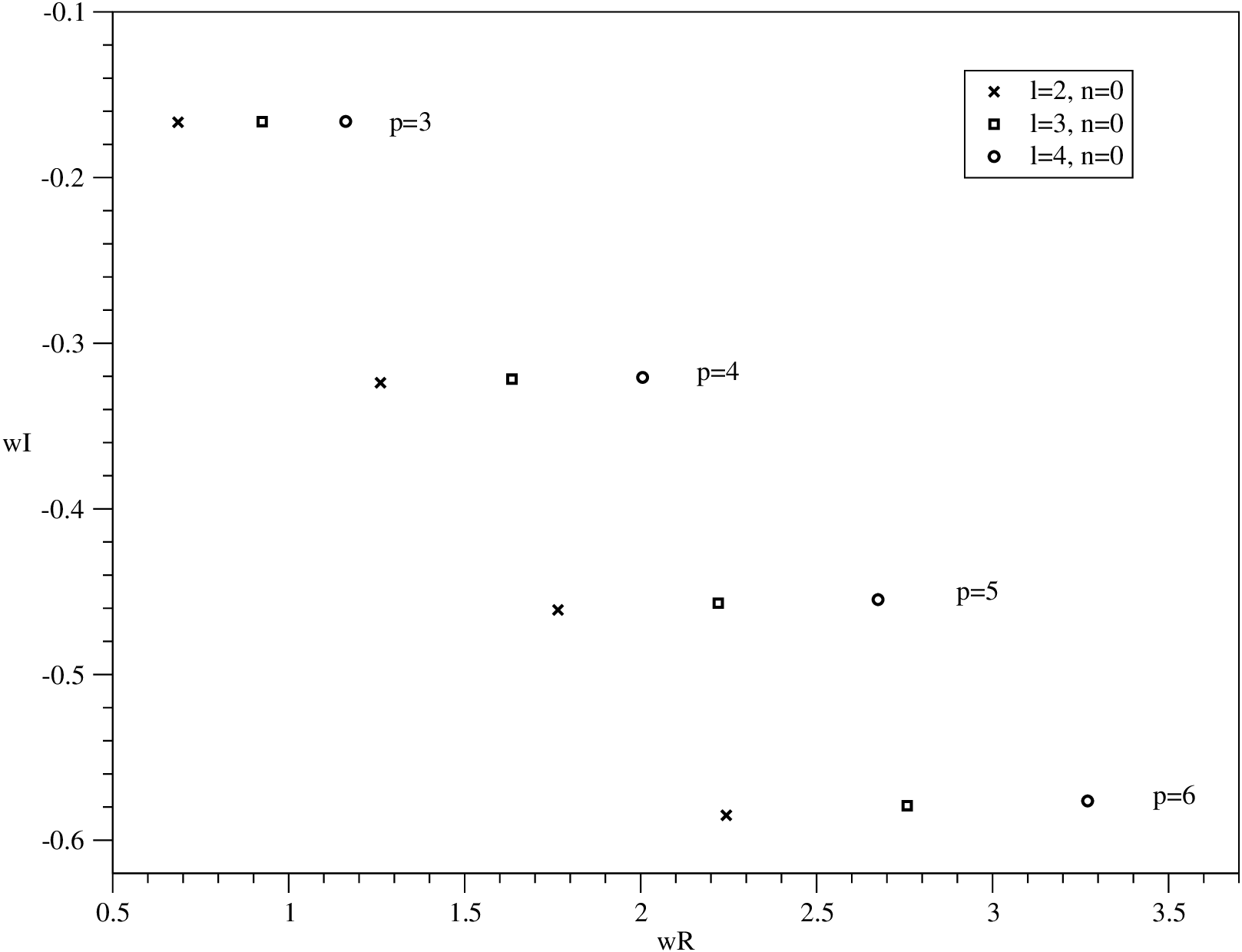}
\label{fig: vector dimension}
\end{center}
\end{figure}

\begin{figure}[htbp]
\begin{center}
\caption{Dependence on the dimension of the QN frequencies for the \textbf{scalar type} electromagnetic perturbations moving in SdS black hole with $\lambda=.070$, for $n=0$ and $l=2$  (crosses), $l=3$ (squares), and $l=4$ (circles).}
\psfrag{wI}{$\omega_I^s$}
\psfrag{wR}{$\omega_R^s$}
\includegraphics[clip,scale=.43]{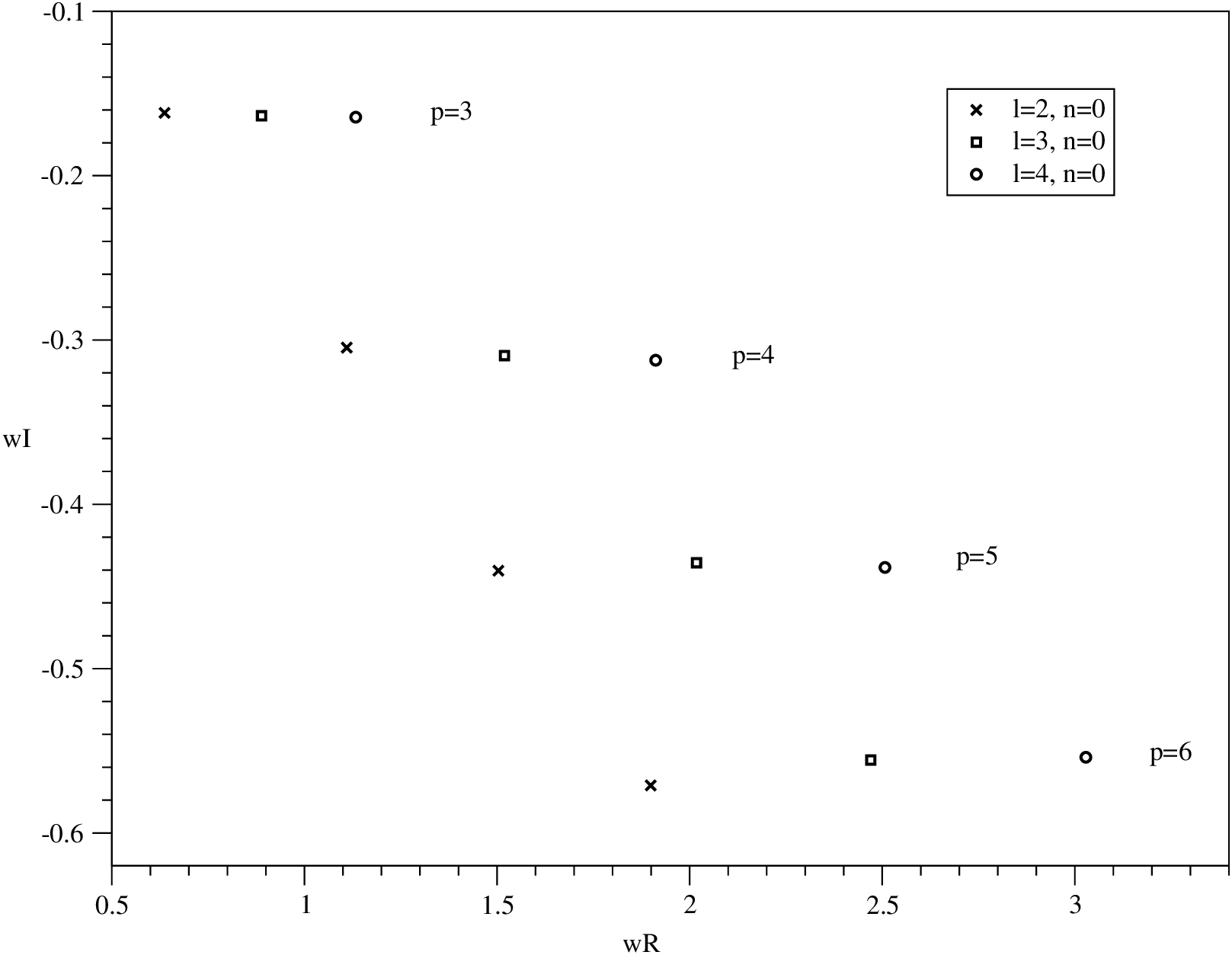}
\label{fig: scalar dimension}
\end{center}
\end{figure}

For $\lambda=.070$ and for vector type and scalar type electromagnetic perturbations, in Figs.\ \ref{fig: vector dimension} and \ref{fig: scalar dimension} we plot the dependence of the QN frequencies on the spacetime dimension. In these figures we observe that for fixed $l$ and $n$ the imaginary part decreases and the real part increases as the number of dimensions increases similar to Schwarzschild black hole previously discussed in Sect.\ \ref{Section Schwarzschild}. 

For the vector type and scalar type electromagnetic perturbations and for fixed $p$, the imaginary parts of the QN frequencies are almost equal for $l=2,3,4$, and $n=0$ (in the scale of the figure), while for different values of $l$ the real parts of the QN frequencies are distinct, (the real parts of the QN frequencies increases as the angular momentum number $l$ increases). These facts agree with formula (\ref{e: SdS large l}) (see the following subsection), from where we can see that in the large angular momentum limit the real part of each mode is proportional to $l$ and the imaginary part tends to a constant value which depends on the spacetime dimension and the mode number. It is convenient to note that formula (\ref{e: SdS large l}) is obtained in the limit $l\to\infty$, but as is well known it also works well for small values of the multipole number \cite{Cho:2007zi}, \cite{Kanti:2005xa}.

From Fig.\ 1 of Ref.\ \cite{Konoplya:2003dd} and our Fig.\ \ref{fig: Re-sds-vector-l=3} we deduce that for SdS black hole the real parts of the QN frequencies for the gravitational $\omega_{R,G}^v$ and electromagnetic perturbations satisfy
\begin{equation} \label{e: QN electromagnetic gravitational}
\omega^v_R > \omega_{R,G}^v,
\end{equation} 
for $l=3$, $n=0$, and $p=3,4,5$. From Fig.\ 2 of Ref.\ \cite{Konoplya:2003dd} and our results we cannot infer the inequality that satisfy the imaginary parts of the QN frequencies for the gravitational and electromagnetic perturbations of vector type. 

Nevertheless our preliminary calculations for the QN frequencies of the vector type gravitational perturbations show that for $p=3,4,5,6$, $n=0$, and $l=2,3,4,5$ their real parts satisfy the inequality (\ref{e: QN electromagnetic gravitational}), while for some values of $l$, $p$, and $\lambda$ the imaginary parts of the frequencies for the vector type electromagnetic field are greater than those of the vector type gravitational perturbations and for other values of $l$, $p$, and $\lambda$ those of the gravitational perturbations are greater.

Furthermore, see Tables V and VI of the recently published work \cite{Konoplya:2007jv}.\footnote{ The QN frequencies provided in Table VI of Ref.\ \cite{Konoplya:2007jv} correspond to gravitational perturbations of vector type.} In these tables the fundamental QN frequencies for $l=2$ of the electromagnetic (Table V) and gravitational (Table VI) perturbations of vector type moving in $D$-dimensional SdS black holes are found for $D=5,6,7,8,9,10,$ and several values of the cosmological constant. We note that for the same values of $D$ and the cosmological constant, the imaginary parts of the QN frequencies for the gravitational perturbations may be greater or less than the corresponding quantities of the electromagnetic fields. We believe that the implications of this fact deserve further investigation, because it is different from the previously known result for Schwarzschild black hole.

Moreover in Ref.\ \cite{Konoplya:2007jv} for $D$-dimensional SdS spacetime the QN frequencies of the scalar type electromagnetic perturbations were not calculated. We believe that additional research on this type of electromagnetic perturbation is necessary, in particular it is interesting to investigate whether the imaginary parts of their QN frequencies are always greater than corresponding quantities of the gravitational perturbations or if these can be smaller, as for vector type perturbations.

For near extremal $D$-dimensional SdS black hole, the QN frequencies of the electromagnetic field were computed analytically in Appendix B of Ref.\ \cite{Lopez-Ortega:2006my}. In that paper was noted that for near extremal $D$-dimensional SdS black hole the effective potentials (\ref{e: SdS vector}) and (\ref{e: SdS scalar}) take the form of a P\"oschl-Teller potential and therefore by using the results of Refs.\ \cite{Mashhoon:1984yo} we can calculate exactly their QN frequencies. For more details see Appendix B of \cite{Lopez-Ortega:2006my} and references cited therein. Our numerical values are not sufficiently accurate for $\lambda$ near the extremal value (\ref{e: SdS black hole condition}); therefore we cannot compare our findings with those obtained by employing the analytical formulas of Ref.\ \cite{Lopez-Ortega:2006my} for near extremal $D$-dimensional SdS black hole.

As for $D$-dimensional Schwarzschild black hole (Sect.\ \ref{Section Schwarzschild}) we do not compare our results for the QN frequencies of the electromagnetic field propagating in SdS backgrounds with those provided in Table V of Ref.\ \cite{Kanti:2005xa} for the same field propagating in Brane World SdS black hole, since in this last reference the fields are evolving in a spacetime that is effectively four dimensional, and the extra spatial dimensions only modify this metric in four dimensions.

We finally point out that for $p=2$, in Schwarzschild black hole both potentials (\ref{eq: potential vector general}) and (\ref{eq: potential scalar general}) are equal and therefore these yield the same QN frequencies for the electromagnetic field propagating in this background. This also happens for SdS black hole in four dimensions.

\subsection{Large angular momentum limit}
\label{Subsection Large l}

As is well known, in the large angular momentum limit ($l \to \infty$) it is possible to find an analytical expression for the QN frequencies of several black holes \cite{Konoplya:2003ii}, \cite{Cho:2007zi}, \cite{Vanzo:2004fy}, \cite{Iyer:1986np}, \cite{Zhidenko:2003wq}, \cite{Press:1971}. In this subsection we calculate the large angular momentum limit of the QN frequencies for the electromagnetic field propagating in $D$-dimensional SdS black hole. 

Our result (\ref{e: SdS large l}) has been previously obtained in Ref.\ \cite{Vanzo:2004fy} by Vanzo and Zerbini using the analytic dilatation approach. For completeness, here by exploiting the first order WKB approximation we provide an alternative derivation.

As $l \to \infty$, both effective potentials (\ref{e: SdS vector}) and (\ref{e: SdS scalar}) simplify to the expression
\begin{equation} \label{e: potential limit large l}
 V(r) = f(r) \frac{l(l+p-1)}{r^2},
\end{equation} 
and using the first order approximation of WKB formula (\ref{e: WKB frequencies}) we find that the QN frequencies are determined by \cite{Schutz-Will:1985}, \cite{Iyer:1986np}
\begin{equation} \label{e: WKB first order}
 \omega^2 = V_0 - i(n+\tfrac{1}{2})(-2 V_0^{\prime \prime} )^{1/2},
\end{equation} 
where $V_0$ and $V_0^{\prime \prime}$ were defined in Sect.\ \ref{Section WKB}.

The effective potential (\ref{e: potential limit large l}) has its maximum at 
\begin{equation}
 r_{\textrm{max}} = (p+1)^{1/(p-1)},
\end{equation} 
and as a consequence we find that the quantities $V_0$ and $V_0^{\prime \prime}$ are equal to
\begin{align} \label{e: potential maximum}
V_0 &= \frac{l(l+p-1)}{(p+1)^{(p+1)/(p-1)}}(p-1-\lambda (p+1)^{(p+1)/(p-1)} ) ,\\ \label{e: potential second maximum}
V_0^{\prime \prime} &= - \frac{2l(l+p-1)(p-1)}{(p+1)^{2(p+1)/(p-1)}}(p-1-\lambda (p+1)^{(p+1)/(p-1)})^2.
\end{align} 
Substituting expressions (\ref{e: potential maximum}) and (\ref{e: potential second maximum}) into formula (\ref{e: WKB first order}), we find that in the limit $l \to \infty$ the QN frequencies are given by
\begin{equation} \label{e: SdS large l}
\omega = \frac{(p-1-\lambda (p+1)^{(p+1)/(p-1)} )^{\tfrac{1}{2}} }{(p+1)^{(p+1)/2(p-1)}}\left[l + \tfrac{p-1}{2} - i\left(n + \tfrac{1}{2} \right) (p-1)^{\tfrac{1}{2}} \right],
\end{equation} 
which is the result already obtained by Vanzo and Zerbini in \cite{Vanzo:2004fy} (see Eqs.\ (5.7) and (7.4) in that reference). Only notice that in Ref.\ \cite{Vanzo:2004fy} the mass of SdS spacetime appear explicitly in the final expressions. In this section it has a fixed value ($M=1$).

Formula (\ref{e: SdS large l}) generalizes to higher dimensions the expression obtained by Zhidenko in expression (19) of Ref.\ \cite{Zhidenko:2003wq}, and simplifies to Eqs.\ (12) and (13) of Ref.\ \cite{Konoplya:2003ii} in the Schwarzschild limit. (See also Eq.\ (38) of Ref.\ \cite{Cho:2007zi} and the comments in Ref.\ \cite{Konoplya:2007jv}.) In formula (19) of Ref.\ \cite{Zhidenko:2003wq} the mass appear explicitly. Konoplya in Ref.\ \cite{Konoplya:2003ii} measured the mass in different units and therefore the Schwarzschild limit of our result (\ref{e: SdS large l}) must be multiplied by a factor of $1/2^{(D-4)/(D-3)}$ to get that of Ref.\ \cite{Konoplya:2003ii}.

Also, Berti, et al.\ in Ref.\ \cite{Berti:2003si} computed the same limit for the QN frequencies of the gravitational perturbations moving in $D$-dimensional Schwarzschild black hole (see formula (26) in \cite{Berti:2003si}). In the previous reference a different normalization for the mass is used. (The Schwarzschild limit of our result (\ref{e: SdS large l}) must be multiplied by $2^{1/(p-1)}$ to get formula (26) of Ref.\ \cite{Berti:2003si}.) 

It is convenient to note that Konoplya in Ref.\ \cite{Konoplya:2003ii} calculated the large angular momentum limit for the QN frequencies of the Klein-Gordon field, but in this limit the effective potentials for the electromagnetic, gravitational, and Klein-Gordon perturbations simplify to the form (\ref{e: potential limit large l}), and therefore the QN frequencies of these three fields are equal as $l \to \infty$, thus they are isospectral in this limit.

\section{$D$-dimensional massless topological black hole}
\label{Section Birmingham}

The knowledge of many exactly solvable systems is useful and allow us to study in detail some properties of the physical phenomena; moreover in other cases the exactly solvable systems are limits of more realistic problems and can be used as test beds for new methods of computation before we apply these methods to study more realistic and complicated questions. 

In $D$-dimensional spacetimes ($D>4$) we know some fields for which an exact computation of their QNMs is possible, see for example \cite{Birmingham:2006zx}--\cite{Lopez-Ortega:2006my} and references therein. We can expect that the exactly calculated QN frequencies may play an important role in future research. In this section we provide an additional example.\footnote{See also Appendix \ref{Appendix Nariai}.}

A simple solution of the Einstein equations with negative cosmological constant is the $D$-dimensional massless topological black hole \cite{Aros:2002te}, \cite{Vanzo:1997gw}, whose metric can take the form (\ref{e: metric general}), where for this black hole the quantity ${\rm d} \sigma^2_p$, is the line element of a $(D-2)$-dimensional compact Einstein space of negative curvature and the function $f(r)$ is equal to
\begin{equation} \label{e: topological black hole}
f(r)= -1+\frac{r^2}{L^2},
\end{equation} 
where the quantity $L$ is related with the cosmological constant. According to Birmingham and Mokhtari this metric can be considered a higher dimensional analogue of BTZ black hole \cite{Birmingham:2006zx}.

The QNMs of the massless topological black hole (\ref{e: topological black hole}) are solutions to the equations of motion for the fields that satisfy the boundary conditions a) the field is ingoing at the event horizon, b) the field is equal to zero at infinity \cite{Aros:2002te}. Recently the QN frequencies for a massive scalar field \cite{Aros:2002te} and for the three types of gravitational perturbations \cite{Birmingham:2006zx} propagating in this spacetime were calculated exactly. Here we compute the corresponding QN frequencies for the two types of electromagnetic perturbations. Also see \cite{Koutsoumbas:2006xj} for other references in which the propagation of fields in several topological black holes is studied.

We first note that in a $D$-dimensional massless topological black hole the effective potentials (\ref{eq: potential vector general}) and (\ref{eq: potential scalar general}) of the electromagnetic field take the form\footnote{In order to exploit the results of Ref.\ \cite{Birmingham:2006zx}, for the spacetime dimension we only use the quantity $D$ to write all the formulas of this section.}
\begin{equation} \label{e: potential vector em}
V_V(r) = \frac{f(r)}{r^2}\left( Q_V - \frac{(D-2)(D-4)}{4} + \frac{(D-2)(D-4)}{4} \frac{r^2}{L^2} \right),
\end{equation} 
for vector type modes and
\begin{equation}\label{e: potential scalar em}
V_S(r) = \frac{f(r)}{r^2}\left( Q_S - \frac{(D-2)(D-4)}{4} + \frac{(D-4)(D-6)}{4} \frac{r^2}{L^2} \right),
\end{equation} 
for scalar type modes. In the previous two equations $Q_V = k_V^2-1$ and $Q_S=k_S^2$ where $k_V^2$ and $k_S^2$ are the eigenvalues of the Laplacian on the $(D-2)$-dimensional compact Einstein space of negative curvature for vector type and scalar type tensor harmonics .

The effective potentials (\ref{e: potential vector em}) and (\ref{e: potential scalar em}) are equal to the effective potentials of the vector type and scalar type gravitational perturbations, respectively. Therefore using the results by Birmingham and Mokhtari of Ref.\ \cite{Birmingham:2006zx} we get the following QN frequencies for the vector type electromagnetic perturbations
\begin{align} \label{e: topological QN vector}
4D, &\qquad \omega_V = \pm \frac{\xi_V}{L} - \frac{2 i}{L} \left( n + \frac{3}{4} \right), \nonumber \\
5D, &\qquad \omega_V = \pm \frac{\xi_V}{L} - \frac{2 i}{L} \left( n + 1 \right),  \\
D\geq 6, &\qquad  \omega_V = \pm \frac{\xi_V}{L} - \frac{2 i}{L} \left( n + \frac{D-1}{4} \right), \nonumber
\end{align}
where the mode number takes the values $n=0,1,2,...$, and for the scalar type electromagnetic perturbations we find
\begin{align} \label{e: topological QN scalar}
4D, &\qquad \omega_S = \pm \frac{\xi_S}{L} - \frac{2 i}{L} \left( n + \frac{3}{4} \right), \nonumber \\
5D, &\qquad \omega_S = \pm \frac{\xi_S}{L} - \frac{2 i}{L} \left( n + \frac{1}{2} \right),  \\
D\geq 6, &\qquad  \omega_S = \pm \frac{\xi_S}{L} - \frac{2 i}{L} \left( n + \frac{D-3}{4} \right). \nonumber
\end{align}
In the preceding two formulas (\ref{e: topological QN vector}) and (\ref{e: topological QN scalar}), for each $D$ we use the quantities $\xi_V$ and $\xi_S$ defined by the following expressions
\begin{equation}
\xi_V^2 = Q_V - \left( \frac{D-3}{2}\right)^2, \qquad \xi_S^2 = Q_S - \left( \frac{D-3}{2}\right)^2.
\end{equation} 

Thus we get an identical mathematical expression for the QN frequencies of the vector type gravitational and electromagnetic perturbations of the $D$-dimensional massless topological black hole. This also occurs for the scalar type gravitational and electromagnetic perturbations. Something similar happens for the QN frequencies of de Sitter spacetime, as was shown in Ref.\ \cite{Lopez-Ortega:2006my}. We only notice the following detail, for the electromagnetic and gravitational perturbations the considered eigenvalues of the Laplacian in each case can be different.

To finish the present section we point out that the effective potentials (\ref{e: potential vector em}) and (\ref{e: potential scalar em}) can take the P\"oschl-Teller form
\begin{equation} \label{e: topological tortoise}
V_{V,S}(x) = \frac{A_{V,S}}{\cosh^2(x/L)} + \frac{B_{V,S}}{\sinh^2(x/L)} ,
\end{equation} 
where $x$ is the tortoise coordinate (\ref{e: tortoise coordinate}), which for the massless topological black hole (\ref{e: topological black hole}) is equal to
\begin{equation}
 x = - L \coth^{-1} \left( \frac{r}{L} \right),
\end{equation} 
and satisfies $x \to - \infty$ as $r \to L$ and $x \to 0$ as $r \to \infty$.

The constants $A_{V,S}$ and $B_{V,S}$ for both gravitational and electromagnetic perturbations of vector type and scalar type take the form
\begin{align}
A_V &=  \frac{1}{L^2} \left( Q_V - \frac{(D-2)(D-4)}{4}\right), \quad \quad B_V=  \frac{(D-2)(D-4)}{4L^2}, \nonumber \\
A_S &=  \frac{1}{L^2} \left( Q_S - \frac{(D-2)(D-4)}{4} \right), \quad \quad B_S= \frac{(D-4)(D-6)}{4L^2} .
\end{align}
For other examples of spacetimes whose effective potentials can take the form (\ref{e: topological tortoise}), see Table 1 in Appendix C of Ref.\ \cite{Lopez-Ortega:2006my}, where a comprehensive list is provided.

\section{Discussion}
\label{Discussion section}

Here for the electromagnetic field propagating in $D$-dimensional Schwarzschild and SdS black holes we compute its low mode number QN frequencies, filling a gap in the literature. In Ref.\ \cite{Lopez-Ortega:2006vn} we show that in these spacetimes the asymptotic behavior of the QN frequencies for the electromagnetic perturbations is different from that for the gravitational perturbations. Nevertheless, in some cases the low mode number QN frequencies of the electromagnetic field behave similarly to those of the gravitational perturbations (or other massless fields) propagating in the same background.

For example, the low mode number QN frequencies of the electromagnetic field moving in Schwarzschild black hole with $D=5,6,7,8,9,10,$ have a similar behavior to those of the gravitational perturbations propagating in this spacetime. For identical values of $p$, $l$, and $n$ the vector type electromagnetic perturbations are more damped and have a greater oscillation frequency than the scalar type electromagnetic fields, except for $p=4$, $l=n=1$, as we noted in Sect.\ \ref{Section Schwarzschild}. Notice too that for the same type of perturbation (scalar or vector) and the same values of $p$, $l$, and $n$ the QN frequencies of the electromagnetic field are more damped and their real parts are greater than those corresponding to gravitational perturbations.

Using different methods, for example those exploited in Ref.\ \cite{Konoplya:2007jv}, for Schwarzschild black hole we should compute the values of the QN frequencies for scalar type electromagnetic fields distinguished with an asterisk in Tables \ref{tab: Schwarzschild p=5}--\ref{tab: Schwarzschild p=8} to prove or disprove these results. For these values of the quantities $p$ and $l$, outside the horizon the effective potentials in addition to the usual maximum, they have a minimum near the horizon and due to the existence of a secondary scattering process near this minimum, we expect that the WKB formula (\ref{e: WKB frequencies}) is not valid. Nevertheless we find that the third, fourth, fifth, and sixth order WKB formulas converge to the values given in Tables \ref{tab: Schwarzschild p=5}--\ref{tab: Schwarzschild p=8}. Thus the new values for these QN frequencies may be useful to determine if we can use the WKB approximation despite the existence of the additional minimum in the effective potential or it is necessary an improvement of this method in order to use it in similar cases, for example, for the electromagnetic perturbations of scalar type with $l=1$ moving in seven- and eight-dimensional SdS black holes (see Sect.\ \ref{Section SdS}).

For $D$-dimensional SdS black hole, as the cosmological constant increases from zero up to the extremal value, the real and imaginary parts of the QN frequencies for the electromagnetic field behave in a similar way to the real and imaginary parts of the QN frequencies for the gravitational perturbations and other massless fields previously computed in Refs.\ \cite{Kanti:2005xa}, \cite{Konoplya:2003dd}, \cite{Zhidenko:2003wq}--\cite{Otsuki:2004yo}. 

Moreover, in $D$-dimensional SdS black hole for some values of the cosmological constant, multipole number, and mode number we have found evidence that for  the QN frequencies of the vector type electromagnetic field are less damped than the corresponding QN frequencies of the vector type gravitational perturbation. Thus the cosmological horizon of SdS black hole change the behavior of the fields observed in asymptotically flat black holes. If something similar happens for scalar type electromagnetic and gravitational perturbations is an interesting question. 

In SdS black hole, for $l=1$ we found some values of $\lambda$ and $D$ for which the scalar type electromagnetic fields are more damped than those of vector type. This behavior is different from that observed in Schwarzschild black hole. For Schwarzschild spacetime the vector type electromagnetic perturbations are more damped than those of scalar type. We believe that the implications of these results deserve further investigation.

As far as we know the numerical calculation of the asymptotic QN frequencies of the electromagnetic field propagating in Schwarzschild ($D>5$) and SdS ($D\geq5$) black holes is lacking. It would be interesting to calculate numerically these asymptotic QN frequencies to prove or disprove the results of Ref.\ \cite{Lopez-Ortega:2006vn}.

For some simple solutions of the Einstein equations, as the massless topological black hole (see Sect.\ \ref{Section Birmingham}) and de Sitter spacetime (see Ref.\ \cite{Lopez-Ortega:2006my}), the QN frequencies of the vector type electromagnetic and gravitational perturbations are determined by the same mathematical expression. This also happens for the scalar type electromagnetic and gravitational perturbations. Therefore only the type of the perturbation (vector or scalar) determines the mathematical form of the QN frequencies corresponding to these backgrounds. Notice that in these spacetimes the gravitational perturbations of tensor type have different QN frequencies than the vector type and scalar type perturbations.\footnote{Moreover the QN frequencies of the tensor type gravitational perturbations are equal to those of the massless Klein-Gordon field.}

\section{Acknowledgments}

This work was supported by CONACyT and SNI (M\'exico).

\begin{appendix}

\section{QN frequencies of $D$-dimensional Nariai solution}
\label{Appendix Nariai}

As we previously commented, there are $D$-dimensional spacetimes whose QN frequencies can be exactly calculated (see for example Refs.\ \cite{Birmingham:2006zx}--\cite{Lopez-Ortega:2006my}, and the previous Sect.\ \ref{Section Birmingham}). One of these spacetimes is the $D$-dimensional Nariai background \cite{b: Nariai solution} for which the QN frequencies of a massless Klein-Gordon field were computed exactly in Ref.\ \cite{Vanzo:2004fy} (and therefore those of the tensor type gravitational perturbation). Moreover the QN frequencies of a massive Klein-Gordon field were calculated in Appendix C of Ref.\ \cite{Lopez-Ortega:2006my}. 

In this appendix our main objective is to compute the QN frequencies of the coupled electromagnetic and gravitational perturbations of scalar type and vector type propagating in $D$-dimensional charged Nariai spacetime \cite{Kodama:2003kk}. So we provide additional perturbations for which their QN frequencies can be exactly calculated in this background and therefore we extend the results of Refs.\ \cite{Vanzo:2004fy}, \cite{Lopez-Ortega:2006my}. Note too that there are no coupled electromagnetic and gravitational perturbations of tensor type \cite{Kodama:2003kk}.

The metric of $D$-dimensional charged Nariai spacetime is given by \cite{Kodama:2003kk}
\begin{equation} \label{e: metric Nariai}
{\rm d} s^2 = - f(\rho)\, {\rm d} t^2 + \frac{ {\rm d} \rho^2}{f(\rho)} + a^2 \,{\rm d} \sigma_p^2 ,
\end{equation} 
where 
\begin{align}
 f(\rho) &= 1 - \sigma \rho^2, \qquad \qquad 
\sigma = (p+1) \lambda - \frac{(p-1)^2 Q^2}{a^{2 p}},
\end{align} 
the quantity $Q$ denotes the electric charge of the spacetime and $a$ is a solution to the equation
\begin{equation}
 (p-1)\frac{K}{a^2} = (p+1)\lambda + \frac{(p-1)Q^2}{a^{2p}}.
\end{equation} 
Here we only study the solution (\ref{e: metric Nariai}) for $\sigma > 0$ and notice that if we satisfy this condition then there are horizons at $\rho = \pm 1/\sqrt{\sigma}$.

The QNMs of Nariai spacetime are solutions to the equations of motion that are purely outgoing near both horizons \cite{Vanzo:2004fy}. For the coupled electromagnetic and gravitational perturbations of vector type propagating in $D$-dimensional Nariai spacetime (\ref{e: metric Nariai}), in Ref.\ \cite{Kodama:2003kk} Kodama and Ishibashi showed that the equations of motion can be simplified to the decoupled partial differential equations (see Eqs.\ (4.43) of Ref.\ \cite{Kodama:2003kk})
\begin{equation} \label{e: vector wave equation}
-\partial_t^2 \Phi_{\pm}^V + f(\rho) \partial_\rho ( f(\rho) \partial_\rho \Phi_{\pm}^V ) - f(\rho) \left(\frac{ k_V^2 + (p-1)K }{a^2} - \sigma \pm \Delta_N  \right) \Phi_{\pm}^V = 0,
\end{equation} 
where 
\begin{equation}
 \Delta_N = \left[ \sigma^2 + \frac{2p(p-1)Q^2}{a^{2p+2}}(k_V^2 + (p-1)K) \right]^{1/2},
\end{equation} 
and $ \Phi_\pm^V $ are the master variables (see Ref.\ \cite{Kodama:2003kk} for more details).

If the quantities $\Phi_\pm^V$ depend on time as
\begin{equation}
 \Phi_\pm^V = \textrm{e}^{-i \omega t} \phi_\pm^V ,
\end{equation} 
then Eqs.\ (\ref{e: vector wave equation}) transform into the ordinary differential equations
\begin{equation} \label{e: Schrodinger radial Nariai}
\left[ f(\rho) \frac{{\rm d}}{{\rm d}\rho}\left( f(\rho)  \frac{{\rm d}}{{\rm d}\rho} \right) + \omega^2 - U_\pm^{(V)} f(\rho) \right] \phi_\pm^V = 0,
\end{equation} 
where the quantities $U_\pm^{(V)}$ are equal to
\begin{equation}
U_\pm^{(V)} = \frac{k_V^2 + (p-1)K}{a^2} - \sigma \pm \left[ \sigma^2 + \frac{2p(p-1)Q^2}{a^{2p+2}}(k_V^2 + (p-1)K) \right]^{1/2}.
\end{equation} 

Using the definition of the tortoise coordinate (\ref{e: tortoise coordinate}), we get that for Nariai spacetime it is given by the following expression
\begin{equation} \label{e: tortoise Nariai}
 x = \int \frac{{\rm d} \rho}{f(\rho)} = \frac{1}{\sigma^{1/2}} \tanh^{-1}(\sigma^{1/2} \rho),
\end{equation} 
and satisfies $x \to \pm \infty$ as $\rho \to \pm 1/\sigma^{1/2}$. Hence, using the tortoise coordinate (\ref{e: tortoise Nariai}) we can write Eqs.\ (\ref{e: Schrodinger radial Nariai}) in the form
\begin{equation} \label{e: potential Poschl Teller}
\left( \frac{{\rm d}^2}{{\rm d} x^2} + \omega^2 - \frac{U_\pm^{(V)}}{\cosh^{2}(\sigma^{1/2} x)} \right) \phi_\pm^V = 0, 
\end{equation} 
that is, as a Schr\"odinger type equation with a P\"oschl-Teller potential.

As is well known the QN frequencies of the P\"oschl-Teller potential which appears in Eq.\ (\ref{e: potential Poschl Teller}) can be exactly computed (see for example Refs.\ \cite{Mashhoon:1984yo}). Therefore using these results we find that the QN frequencies for the coupled electromagnetic and gravitational perturbations of vector type are equal to
\begin{equation} \label{e: quasinormal frequency vector}
\omega_V = \pm \sqrt{ U_\pm^{(V)} - \frac{\sigma}{4} } - i \sigma^{1/2} \left( n + \frac{1}{2} \right).
\end{equation} 

Moreover, for the coupled electromagnetic and gravitational perturbations of scalar type propagating in $D$-dimensional Nariai spacetime, in Ref.\ \cite{Kodama:2003kk} Kodama and Ishibashi showed that their equations of motion can be simplified to the ordinary differential equations for the master variables $\phi_\pm^S$ (see Eqs.\ (5.77) of Ref.\ \cite{Kodama:2003kk})
\begin{equation} \label{e: scalar Schrodinger type}
\left[  f(\rho) \frac{{\rm d}}{{\rm d}\rho}\left( f(\rho)  \frac{{\rm d}}{{\rm d}\rho} \right) + \omega^2  - \left( \frac{k_S^2}{a^2} - \sigma \pm \mu  \right) \right] \phi_\pm^S = 0,
\end{equation} 
where
\begin{equation}
\mu = \left( \sigma^2 + \frac{4 (p-1)^2 Q^2}{a^{2p + 2}} k^2_S \right)^{1/2}.
\end{equation} 

Using the tortoise coordinate $x$ of expression (\ref{e: tortoise Nariai}) we can transform Eqs.\ (\ref{e: scalar Schrodinger type}) into
\begin{equation}
\left( \frac{{\rm d}^2}{{\rm d} x^2} + \omega^2 - \frac{U_\pm^{(S)}}{\cosh^{2}(\sigma^{1/2} x)} \right) \phi_\pm^S  = 0,
\end{equation} 
where
\begin{equation} 
U_\pm^{(S)} = \frac{k_S^2}{a^2} - \sigma \pm \left( \sigma^2 + \frac{4 (p-1)^2 Q^2}{a^{2p + 2}} k^2_S \right)^{1/2}.
\end{equation} 
Thus if we replace the constants $U_\pm^{(V)}$ by $U_\pm^{(S)}$, then the expression (\ref{e: quasinormal frequency vector}) also gives the QN frequencies for this type of coupled perturbations.

Therefore we find that for both types of coupled electromagnetic and gravitational perturbations the QN frequencies are determined by expression (\ref{e: quasinormal frequency vector}). To our knowledge these results represent the first exact calculation of the QN frequencies for the coupled electromagnetic and gravitational perturbations in higher dimensions.

To finish this appendix we note that there are a few analytical results for the QN frequencies of the coupled electromagnetic and gravitational perturbations propagating in charged backgrounds. See Ref.\ \cite{Molina:2003dc}, for the analytically computed QN frequencies of the coupled perturbations moving in a four-dimensional near extremal Reissner-Nordstr\"om de Sitter black hole. Nevertheless the previous comparisons with numerical values show that the approximations used in this reference are only valid when the imaginary parts of the QN frequencies are small. Moreover, in Refs.\ \cite{Motl:2003cd}--\cite{Andersson:2003fh}, are calculated analytically the asymptotic QN frequencies for the coupled electromagnetic and gravitational perturbations in four and $D$-dimensional Reissner-Nordstr\"om black holes.

\end{appendix}

\newpage


\begin{table}
\centering
\caption{QN frequencies of the vector type and scalar type electromagnetic perturbations moving in a five-dimensional Schwarzschild black hole (\mbox{\boldmath $p=3$}).}
\begin{tabular}[htp]{llllll}
\hline
$l$ & $n$ & $\omega_R^v$ & $\omega_I^v$ & $\omega_R^s $ & $\omega_I^s $ \\ \hline
1 & 0 & 0.6728 & -0.2496 & 0.5210 &	-0.2230 \\ \hline
  & 1 &	0.5540 & -0.8025 & 0.3402 &	-0.7414 \\ \hline
2 & 0 &	1.0384 & -0.2492 & 0.9484 &	-0.2385 \\ \hline
  & 1 &	0.9534 & -0.7706 & 0.8566 &	-0.7381 \\ \hline
  & 2 &	0.8039 & -1.3604 & 0.6908 &	-1.3097 \\ \hline
3 & 0 &	1.3976 & -0.2496 & 1.3306 &	-0.2437 \\ \hline
  & 1 &	1.3326 & -0.7614 & 1.2632 &     -0.7439 \\ \hline
  & 2 &	1.2108 & -1.3121 & 1.1359 &	-1.2839 \\ \hline
  & 3 & 1.0492 & -1.9258 & 0.9654 &     -1.8898 \\ \hline
4 & 0 &	1.7545 & -0.2497 & 1.7011 &	-0.2460 \\ \hline
  & 1 &	1.7021 & -0.7573 & 1.6475 &	-0.7461 \\ \hline
  & 2 &	1.6012 & -1.2896 & 1.5441 &	-1.2713 \\ \hline
  & 3 &	1.4608 & -1.8630 & 1.3995 & 	-1.8386 \\ \hline
  & 4 &	1.2930 & -2.4922 & 1.2260 &	-2.4634 \\ \hline
5 & 0 &	2.1102 & -0.2498 & 2.0659 &	-0.2472 \\ \hline
  & 1 &	2.0664 & -0.7551 & 2.0214 & 	-0.7473 \\ \hline
  & 2 &	1.9809 & -1.2774 & 1.9345 & 	-1.2646 \\ \hline
  & 3 &	1.8587 & -1.8284 & 1.8101 &	-1.8110 \\ \hline
  & 4 &	1.7075 & -2.4191 & 1.6557 &	-2.3977 \\ \hline
  & 5 &	1.5360 & -3.0591 & 1.4801 &	-3.0349 \\ \hline
\end{tabular}
\label{tab: Schwarzschild p=3}
\end{table}


\begin{table}
\centering
\caption{QN frequencies of the vector type and scalar type electromagnetic perturbations moving in a six-dimensional Schwarzschild black hole (\mbox{\boldmath $p=4$}).}
\begin{tabular}[htp]{llllll}
\hline
$l$ & $n$ & $\omega_R^v$ & $\omega_I^v$ & $\omega_R^s $ & $\omega_I^s $ \\ \hline
1 & 0 & 1.1060 & -0.4028 & 0.8169 & -0.4010 \\ \hline
  & 1 & 0.8636 & -1.3148 & 0.5012 & -1.3313 \\ \hline
2 & 0 & 1.5704 & -0.3939 & 1.3591 & -0.3692 \\ \hline
  & 1 & 1.3906 & -1.2248 & 1.1398 & -1.1521 \\ \hline
  & 2 & 1.0460 & -2.2082 & 0.7093 & -2.1082 \\ \hline
3 & 0 & 2.0267 & -0.3933 & 1.8662 & -0.3762 \\ \hline
  & 1 & 1.8862 & -1.2040 & 1.7166 & -1.1508 \\ \hline
  & 2 & 1.6084 & -2.0982 & 1.4150 & -2.0062 \\ \hline
  & 3 & 1.2144 & -3.1541 & 0.9786 & -3.0270 \\ \hline
4 & 0 & 2.4818 & -0.3930 & 2.3506 & -0.3813 \\ \hline
  & 1 & 2.3666 & -1.1948 & 2.2318 & -1.1590 \\ \hline
  & 2 & 2.1369 & -2.0494 & 1.9932 & -1.9883 \\ \hline
  & 3 & 1.8011 & -3.0052 & 1.6410 & -2.9192 \\ \hline
  & 4 & 1.3800 & -4.1180 & 1.1956 & -4.0106 \\ \hline
5 & 0 & 2.9362 & -0.3928 & 2.8252 & -0.3843 \\ \hline
  & 1 & 2.8387 & -1.1897 & 2.7258 & -1.1638 \\ \hline
  & 2 & 2.6436 & -2.0229 & 2.5263 & -1.9792 \\ \hline
  & 3 & 2.3545 & -2.9248 & 2.2292 & -2.8632 \\ \hline
  & 4 & 1.9820 & -3.9335 & 1.8447 & -3.8555 \\ \hline
  & 5 & 1.5440 & -5.0900 & 1.3906 & -4.9983 \\ \hline
\end{tabular}
\label{tab: Schwarzschild p=4}
\end{table}


\begin{table}
\centering
\caption{QN frequencies of the vector type and scalar type electromagnetic perturbations moving in a seven-dimensional Schwarzschild black hole (\mbox{\boldmath $p=5$}). The asterisk indicate that for this value of the multipole number the effective potential is not positive definite outside the horizon.}
\begin{tabular}[htp]{llllll}
\hline
$l$ & $n$ & $\omega_R^v$ & $\omega_I^v$ & $\omega_R^s $ & $\omega_I^s $ \\ \hline
1 & 0 &	1.5345 & -0.5523 & 1.1988$^{*}$ & -0.5217$^{*}$ \\ \hline
  & 1 &	1.1326 & -1.8353 & 0.6891$^{*}$ & -1.7031$^{*}$ \\ \hline
2 & 0 &	2.0784 & -0.5259 & 1.7549 & -0.5052 \\ \hline
  & 1 &	1.7760 & -1.6417 & 1.3805 & -1.5927 \\ \hline
  & 2 & 1.1432 & -3.0187 & 0.6243 & -2.9904 \\ \hline
3 & 0 &	2.6012 & -0.5243 & 2.3440 & -0.4993 \\ \hline
  & 1 & 2.3641 & -1.6047 & 2.0766 & -1.5286 \\ \hline
  & 2 &	1.8615 & -2.8171 & 1.5047 & -2.6932 \\ \hline
  & 3 &	1.1012 & -4.3354 & 0.6393 & -4.1816 \\ \hline
4 & 0 &	3.1238 & -0.5237 & 2.9103 & -0.5035 \\ \hline
  & 1 & 2.9274 & -1.5922 & 2.7036 & -1.5290 \\ \hline
  & 2 & 2.5161 & -2.7407 & 2.2665 & -2.6285 \\ \hline
  & 3 & 1.8780 & -4.0745 & 1.5801 & -3.9096 \\ \hline
  & 4 & 1.0468 & -5.7280 & 0.6789 & -5.5100 \\ \hline
5 & 0 &	3.6464 & -0.5233 & 3.4629 & -0.5075 \\ \hline
  & 1 &	3.4787 & -1.5850 & 3.2909 & -1.5364 \\ \hline
  & 2 & 3.1310 & -2.7011 & 2.9326 & -2.6159 \\ \hline
  & 3 & 2.5891 & -3.9379 & 2.3698 & -3.8116 \\ \hline
  & 4 & 1.8611 & -5.3903 & 1.6086 & -5.2202 \\ \hline
  & 5 & 0.9860 & -7.1556 & 0.6891 & -6.9382 \\ \hline
\end{tabular}
\label{tab: Schwarzschild p=5}
\end{table}


\begin{table}
\centering
\caption{QN frequencies of the vector type and scalar type electromagnetic perturbations moving in an eight-dimensional Schwarzschild black hole (\mbox{\boldmath $p=6$}). The asterisk indicate that for this value of the multipole number the effective potential is not positive definite outside the horizon.}
\begin{tabular}[htp]{llllll}
\hline
$l$ & $n$ & $\omega_R^v$ & $\omega_I^v$ & $\omega_R^s $ & $\omega_I^s $ \\ \hline
1 & 0 & 1.9515 & -0.7038 & 1.5757$^{*}$ & -0.6335$^{*}$ \\ \hline
  & 1 & 1.3631 & -2.3772 & 0.8357$^{*}$ & -2.0502$^{*}$ \\ \hline
2 & 0 & 2.5746 & -0.6462 & 2.1755 & -0.6308 \\ \hline
  & 1 & 2.1208 & -2.0321 & 1.6570 & -1.9586 \\ \hline
  & 2 & 1.1112 & -3.7937 & 0.4886 & -3.6661 \\ \hline
3 & 0 &	3.1465 & -0.6437 & 2.8024 & -0.6192 \\ \hline
  & 1 & 2.7952 & -1.9626 & 2.3976 & -1.8940 \\ \hline
  & 2 & 1.9940 & -3.4524 & 1.4893 & -3.3581 \\ \hline
  & 3 & 0.7166 & -5.4080 & \hspace{.4cm}-- & \hspace{.4cm}-- \\ \hline
4 & 0 & 3.7184 & -0.6435 & 3.4260 & -0.6182 \\ \hline
  & 1 & 3.4254 & -1.9502 & 3.1082 & -1.8730 \\ \hline
  & 2 & 2.7771 & -3.3505 & 2.4027 & -3.2183 \\ \hline
  & 3 &	1.7087 & -5.0298 & 1.2379 & -4.8470 \\ \hline
  & 4 &\hspace{.4cm} -- & \hspace{.4cm}-- & \hspace{.4cm}--     &  \hspace{.4cm}--      \\ \hline
5 & 0 & 4.2913 & -0.6430 & 4.0370 & -0.6210 \\ \hline
  & 1 & 4.0392 & -1.9436 & 3.7745 & -1.8756 \\ \hline
  & 2 & 3.4950 & -3.3056 & 3.2050 & -3.1840 \\ \hline
  & 3 & 2.5978 & -4.8407 & 2.2586 & -4.6561 \\ \hline
  & 4 & 1.3480 & -6.7335 & 0.9329 & -6.4775 \\ \hline
  & 5 &	  \hspace{.4cm}--    &  \hspace{.4cm}--      &  \hspace{.4cm}--     &   \hspace{.4cm}--     \\ \hline
\end{tabular}
\label{tab: Schwarzschild p=6}
\end{table}


\begin{table}
\centering
\caption{QN frequencies of the vector type and scalar type electromagnetic perturbations moving in a nine-dimensional Schwarzschild black hole (\mbox{\boldmath $p=7$}). The asterisk indicate that for this value of the multipole number the effective potential is not positive definite outside the horizon.}
\begin{tabular}[htp]{llllll}
\hline
$l$ & $n$ & $\omega_R^v$ & $\omega_I^v$ & $\omega_R^s $ & $\omega_I^s $ \\ \hline 
1 & 0 & 2.3499 & -0.8640 & 1.9647$^{*}$ & -0.7387$^{*}$ \\ \hline
  & 1 & 1.5654 & -2.9384 & 0.9809$^{*}$ & -2.3294$^{*}$ \\ \hline
2 & 0 & 3.0654 & -0.7567 & 2.6073 & -0.7421 \\ \hline
  & 1 & 2.4254 & -2.4212 & 1.9229 & -2.2603 \\ \hline
  & 2 & 0.9800 & -4.5264 & 0.2127 & -4.1722 \\ \hline
3 & 0 &	3.6754 & -0.7524 & 3.2609 & -0.7329 \\ \hline
  & 1 & 3.1962 & -2.2780 & 2.7170 & -2.2219 \\ \hline
  & 2 & 2.0226 & -3.9927 & 1.4095 & -3.8952 \\ \hline
  & 3 & \hspace{.4cm}-- & \hspace{.4cm}-- &  \hspace{.4cm}--     &   \hspace{.4cm}--     \\ \hline
4 & 0 & 4.2842 & -0.7536 & 3.9214 & -0.7279 \\ \hline
  & 1 & 3.8826 & -2.2687 & 3.4786 & -2.1953 \\ \hline
  & 2 & 2.9416 & -3.8597 & 2.4478 & -3.7453 \\ \hline
  & 3 & 1.2912 & -5.8052 & 0.6549 & -5.6685 \\ \hline
  & 4 &  \hspace{.4cm}--     &  \hspace{.4cm}--      &  \hspace{.4cm}--     &  \hspace{.4cm}--      \\ \hline
5 & 0 & 4.8958 & -0.7535 & 4.5755 & -0.7281 \\ \hline
  & 1 & 4.5475 & -2.2675 & 4.2066 & -2.1902 \\ \hline
  & 2 & 3.7630 & -3.8249 & 3.3737 & -3.6903 \\ \hline
  & 3 & 2.3877 & -5.5843 & 1.9071 & -5.3876 \\ \hline
  & 4 &	\hspace{.4cm}--& \hspace{.4cm}-- &  \hspace{.4cm}--     &  \hspace{.4cm}--      \\ \hline
  & 5 &	 \hspace{.4cm}--     &  \hspace{.4cm}--      &  \hspace{.4cm}--     & \hspace{.4cm} --      \\ \hline
\end{tabular}
\label{tab: Schwarzschild p=7}
\end{table}


\begin{table}
\centering
\caption{QN frequencies of the vector type and scalar type electromagnetic perturbations moving in a ten-dimensional Schwarzschild black hole (\mbox{\boldmath $p=8$}). The asterisk indicate that for this value of the multipole number the effective potential is not positive definite outside the horizon.}
\label{tab: Schwarzschild p=8}
\begin{tabular}[htp]{llllll}
\hline
$l$ & $n$ & $\omega_R^v$ & $\omega_I^v$ & $\omega_R^s $ & $\omega_I^s $ \\ \hline 
1 & 0 & 2.7225 & -1.0397 & 2.3664$^{*}$ & -0.8373$^{*}$ \\ \hline
  & 1 & 1.7563 & -3.4946 & 1.1318$^{*}$ & -2.5324$^{*}$ \\ \hline
2 & 0 &	3.5535 & -0.8590 & 3.0431$^{*}$ & -0.8435$^{*}$ \\ \hline
  & 1 & 2.6832 & -2.8483 & 2.1744$^{*}$ & -2.5113$^{*}$ \\ \hline
  & 2 & 0.7948 & -5.1611 &  \hspace{.4cm}--     &  \hspace{.4cm}--      \\ \hline
3 & 0 & 4.1962 & -0.8509 & 3.7222 & -0.8382 \\ \hline
  & 1 & 3.5779 & -2.5542 & 3.0329 & -2.5029 \\ \hline
  & 2 & 1.9679 & -4.4425 & 1.2311 & -4.2591 \\ \hline
  & 3 &  \hspace{.4cm}--     &  \hspace{.4cm}--      &  \hspace{.4cm}--     &  \hspace{.4cm}--      \\ \hline
4 & 0 & 4.8317 & -0.8550 & 4.4087 & -0.8322 \\ \hline
  & 1 &	4.3144 & -2.5463 & 3.8353 & -2.4881 \\ \hline
  & 2 & 3.0281 & -4.2434 & 2.4260 & -4.1576 \\ \hline
  & 3 & 0.6072 & -6.3179 &  \hspace{.4cm}--     &  \hspace{.4cm}--      \\ \hline
  & 4 &  \hspace{.4cm}--     &  \hspace{.4cm}--      &  \hspace{.4cm}--     &  \hspace{.4cm}--      \\ \hline
5 & 0 & 5.4735 & -0.8562 & 5.0942 & -0.8301 \\ \hline
  & 1 & 5.0202 & -2.5567 & 4.6084 & -2.4817 \\ \hline
  & 2 & 3.9551 & -4.2394 & 3.4672 & -4.1179 \\ \hline
  & 3 & 1.9467 & -6.0904 & 1.3109 & -5.9317 \\ \hline
  & 4 &  \hspace{.4cm}--     &  \hspace{.4cm}--      &  \hspace{.4cm}--     &   \hspace{.4cm}--     \\ \hline
  & 5 &	 \hspace{.4cm}--     &  \hspace{.4cm}--      &  \hspace{.4cm}--     &  \hspace{.4cm}--      \\ \hline
\end{tabular}
\end{table}


\end{document}